\documentclass[a4paper,fleqn,usenatbib]{mnras}

\usepackage{newtxtext,newtxmath}

\usepackage[T1]{fontenc}
\usepackage{ae,aecompl}

\usepackage{graphicx}	
\usepackage{amsmath}	
\usepackage{float}

\newcommand{\rev}[1]{\textcolor{black}{#1}}

\title[Optimal multi-line IM during reionization] 
{Optimal survey parameters: Ly$\alpha$ and H$\alpha$ intensity mapping for synergy with the 21cm signal during reionization}

\author[C. Heneka, A. Cooray]{
Caroline Heneka,$^{1,2}$\thanks{E-mail: caroline.heneka@uni-hamburg.de}
Asantha Cooray$^{3}$
\\
$^{1}$ University of Hamburg, Hamburg Observatory, Gojenbergsweg 112, 21029 Hamburg, Germany \\
$^{2}$ Scuola Normale Superiore, Piazza dei Cavalieri 7, 56126 Pisa, Italy \\
$^{3}$ Department of Physics and Astronomy, University of California, Irvine, CA 92697, USA
}

\date{Accepted XXX. Received YYY; in original form ZZZ}

\pubyear{2021}

\begin{document}
\label{firstpage}
\pagerange{\pageref{firstpage}--\pageref{lastpage}}
\maketitle

\begin{abstract}
Intensity mapping of multiple emission lines is emerging as a new branch to astronomy, to probe both properties of ionizing sources and the medium between, in particular the intergalactic medium. For Epoch of Reionization (EoR) studies, both multi-line experiments and analysis methods are still in their infancy. Here we explore optimal survey parameters for Ly$\alpha$ (and H$\alpha$) intensity mapping up to high redshifts of reionization, and requirements for optimised synergy with 21cm experiments. We investigate line sensitivity, spectral resolution and detector pixel size requirements for optimal (high signal-to-noise) mission output. Power and cross-power spectra in a fiducial setup are derived, as are mock intensity maps. For line power spectrum measurements a cumulative signal-to-noise of O$\left(10^3 \right)$, and for respective cross-spectra with SKA 21cm observations of O$\left(10\right)$ to O$\left(10^2\right)$ are possible per redshift bin around the midpoint of reionization. These high signal-to-noise tomographic measurements are in reach for line sensitivities $>8\times 10^{-18}$erg$\,$s$^{-1}$sr$^{-1}$Hz$^{-1}$, spectral resolution $R>250$ and detector pixel sizes $<2\,$arcsec; all three requirements are met by the proposed Cosmic Dawn Intensity Mapper (CDIM). For CDIM similar S/N values are in reach for H$\alpha$. Already the planned NASA mission SPHEREx will detect during the EoR Ly$\alpha$ autopower and cross power with 21cm (\rev{SKA and HERA-type}), for sensitivities >$10^{-18}$erg$\,$s$^{-1}$sr$^{-1}$Hz$^{-1}$ in a moderate 21cm foreground scenario (>$10^{-17}$erg$\,$s$^{-1}$sr$^{-1}$Hz$^{-1}$ in an optimistic scenario). We advocate for IR missions in flavor of CDIM for a leap in IM and finish by providing a cookbook for successful multi-line IM during the EoR.
\end{abstract}

\begin{keywords}
galaxies: cosmology: dark ages, reionization, first stars -- high-redshift -- intergalactic medium -- diffuse radiation -- infrared: general -- large-scale structure of Universe
\end{keywords}



\section{Introduction}
Intensity  mapping (IM)  of multiple  lines  provides  a  unique  and  new  probe to map out vast cosmological volumes up to high redshifts, revolutionising in the years to come what we can learn about cosmology and astrophysics. Intensity mapping  refers  to  low-resolution  mapping  that  measures  the  entire integrated  emission. It thus  includes the cumulative emission of faint and diffuse sources, alleviating the need of galaxy surveys to resolve individual objects. At the same time spectral lines such as Ly$\alpha$ and H$\alpha$ (including their cosmological redshift information) allow us  to  tomographically map their fluctuations in  three  dimensions. At high redshifts of reionization, the Universe's last big phase transition from neutral in hydrogen to mostly ionised, when line emitting galaxies tend to be too faint to be resolved or produce extremely biased samples, intensity mapping still is able to probe both the ionising sources that drive reionization and the gaseous medium between. 

Complementary to ongoing and upcoming measurements of the cosmological 21cm line signal that probes neutral hydrogen and can thus follow the process of reionization, galactic emission lines at high redshifts are able to probe both independently and jointly with upcoming 21cm experiments the progress of reionization, the properties of ionising sources and the state of the IGM~\citep[e.g.][]{2011ApJ...741...70L,Silva12,Serra:2016jzs, Heneka17,2018revPratika}, besides informing us about underlying cosmology~\citep[e.g.][]{2013JCAP...01..003B,2018JCAP...10..004H,2020JCAP...05..038L}. A wealth of spectral lines such as  CO, CII, OII, NII and H$\alpha$ preferentially probe different aspects of star and galaxy formation, the interstellar medium (ISM) and different phases of the circumgalactic medium (CGM). Especially Ly$\alpha$ emission has been a major observational probe for the high redshift universe, including surveys of Lyman-alpha emitting galaxies~\citep{10.1093/pasj/psx074,2020ARA&A..58..617O}, as due to resonant scattering and recombination Ly$\alpha$ observations witness ionized regions of star formation and AGN activity but also diffuse neutral hydrogen gas in the CGM and the IGM. Current radio interferometers such as the Low Frequency Array~\citep[LOFAR]{2013A&A...556A...2V},\footnote{http://www.lofar.org} the Hydrogen Epoch of Reionization Array~\citep[HERA]{DeBoer:2016tnn},\footnote{http://reionization.org/} and the Murchison Wide-field Array~\citep[MWA]{tingay2013,Beardsley:2016njr},\footnote{http://www.mwatelescope.org} are  aiming  to  statistically  detect  the 21cm signal.  In the near future the Square Kilometre Array (SKA),\footnote{https://skatelescope.org/} will image 21 cm and is expected to detect e.g. a cross-correlation signal with Lyman-alpha emitters (LAEs)  in  just  a  few hours, even under pessimistic assumptions about foreground contamination and the reionization model~\citep{Sobacchi:2016mhx, Hutter:2016,10.1093/mnras/staa183,Kubota20,Heneka2020}. Some examples of line intensity mapping experiments are future CO  experiments such as COMAP~\citep{Li_2016}, and experiments that aim to measure [CII] as TIME~\citet{10.1117/12.2057207} and  CONCERTO~\citep{10.1093/mnras/stz617}. The Spectrophotometer for the History of the Universe, Epoch of Reionization, and Ice Explorer~\citep[SPHEREx]{2014spherex} targets among others Ly$\alpha$ and H$\alpha$ lines and the proposed Cosmic Dawn Intensity Mapper~\citep[CDIM]{2016arXiv160205178C} is able to perform multi-line IM throughout high redshifts of reionization, with an effective telescope aperture size of 83 cm, as detailed in the CDIM science report~\citep{CDIMreport}. 

To both optimally prepare for such experiments and exploit their measurements, the IM signal needs to be consistently connected to the physical properties of line emission as well as cosmological quantities of interest. Empirical modelling has been extensively used in the literature to model the expected emission of different lines, see e.g.~\citet{10.1093/mnras/stv933,Silva_2015,10.1093/mnras/stw2470,Ihle_2019}. Its drawback is the simplistic modelling based on for example luminosity-mass relations alone, without taking into account gas properties and evolution in detail, and thus the difficulty to translate its integrated quantities into meaningful mock observations that include diffuse components. Moreover, empirical approaches often focus on one line without self-consistently modelling several lines alongside each other as is advantageous for multi-line intensity mapping. \rev{Semi-numerical modelling attempts to bring together analytical modelling based on physical parameters of the gaseous medium and for example generalised halo models as in~\citet{Jason2019} for multiple lines.} Hydrodynamical simulations (see for example~\citet{BlueTides15,MCALPINE201672,TNG2019}) model physical processes more robustly, but due to their computational cost, simulating large enough volumes for IM limits resolution attainable and requires phenomenological sub-grid models to include processes such as star formation, while scanning large parameter spaces is computationally unfeasible. Semi-analytical models adopt simplifying assumptions for evolving dark matter and baryons coupled again with phenomenological recipes. They can be difficult to calibrate with observations, and still require orders of magnitude too high computational time for full exploration of both astrophysical and cosmological parameter spaces~\citep{2019MNRAS.487.1946Q,Astra2020,2020arXiv200911933Y}.

Semi-numerical simulations are the tool of choice when modelling large volumes for intensity mapping to explore parameter spaces and provide realistic IM mock observational maps; they have been widely employed for modelling the 21cm IM signal e.g. with 21cm FAST~\citep{Mesinger10,Sobacchi:2014rua}. As compared to N-body/hydro simulations they allow for fast exploration of parameter space to derive constraints, are easy to be calibrated to match and agree with both hydrodynamical simulations and observations on IM-relevant scales, while consistently evolving galactic properties based on a perturbed density field and/or an underlying halo field. We therefore advocate their use to model IM of multiple spectral lines.

In  this  paper  we  seek to explore optimal multi-line IM survey parameters in a semi-numerical setup, suitable to both create realistic mock observational maps and measure autopower and cross-power spectra between rest-frame optical \rev{and UV} lines (H$\alpha$, Ly$\alpha$) during reionization and the famous 21cm line. This is crucial to optimally exploit upcoming measurements and design new IM probes able to map out more than 80 per-cent of the observable Universe in multiple 'colours'. We employ our cosmological volume multi-line emission simulations tailored for IM based on~\citet{Heneka17}, coupled with a fast evaluation module for thermal and instrumental noise, to derive mock observational maps and signal-to-noise response curves attainable for various instrumental setups.

The paper is organised as follows. In section~\ref{sec:sim} we briefly describe our modelling and simulations. In section~\ref{sec:Pk} we present auto and cross-power spectra for 21cm, Ly$\alpha$ and H$\alpha$ lines, followed by the presentation of fiducial mock observations in section~\ref{sec:mocks}. We continue with an exploration of signal-to-noise ratios attainable for different survey parameters, to deduce optimal survey design for multi-line intensity mapping surveys in section~\ref{sec:survey}. We present main findings in our cookbook for multi-line IM in section~\ref{sec:cookbook}, to finish with outlook and conclusions in section~\ref{sec:out}.

\section{Multi-line IM modelling}\label{sec:sim}
Our model for the intensity mapping signal of multiple lines and underlying reionization of the IGM is based on the semi-numerical simulation of the progress of reionization and the corresponding 21cm signal with 21cmFAST and its parent code DexM that includes a halo finder~\citep{DexM07}, coupled with the semi-numerical simulation of galactic, diffuse IGM and scattered IGM emission of a given line consistently evolved alongside. We here summarise the main characteristics of the simulation. For detailed descriptions of the model we refer to~\citep{Heneka17}.

The simulations employed in this work have a box size of 200$\,$Mpc and are computed on a 300$^3$ grid, sufficiently large and resolved to capture proposed multi-line IM survey sizes and accurately model the progress of reionization. Significantly larger box sizes are feasible though, due to the fast computational speed of simulation and mock creation for multiple lines in this framework (less than a few CPU hours for simulating the full multi-line IM evolution during reionization). With density, velocity, ionization and gas kinetic temperature fields, the fluctuations in 21cm brightness temperature offset $\delta \mathrm{T}_\mathrm{b}$ of the spin gas temperature $\mathrm{T}_\mathrm{S}$ from  the  cosmic  microwave  background (CMB) temperature $\mathrm{T}_{\gamma}$ is derived. The simulation self-consistently  computes  the  Lyman  series radiation background, that determines how closely the spin temperature couples to the gas kinetic temperature through Wouthuysen-Field effect~\citep{1952AJ.....57R..31W,1958PIRE...46..240F}. The simulations include inhomogeneous recombinations, photo-heating  suppression in small  halos, and inhomogeneous evolution of IGM temperature pre-reionization calibrated via  X-ray  emissivity  of  galaxies. The reionization model in this work reaches its midpoint of reionization at $z\sim 8$ and its endpoint at $z\sim 6$ (mean IGM neutral fraction $\mathrm{\bar{x}}_\mathrm{HI}<0.01$). 

For modelling the emission of further lines, here Ly$\alpha$ and H$\alpha$, we distinguish between galactic and IGM components. By galactic  component  we  mean  the  contribution from  within  the  virial  radius  of  line-emitting  galaxies themselves. The IGM component comprises a diffuse ionized IGM around galaxies where hydrogen recombines, as well as for the resonant Ly$\alpha$ line the background caused by X-ray/UV heating and down-scattering of Lyman-n photons. 
As galactic emission is closely related to the ionizing photon rate $\dot{N}_\mathrm{ion}$ and therefore star formation, it can be connected to the star formation rate (SFR) of galaxies as $\dot{N}_\mathrm{ion} \propto \mathrm{SFR}$. For the Ly$\alpha$ line, dominant  source of emission  is   hydrogen  recombination,  as  well  as collisional excitation. For both recombination and excitation, the Ly$\alpha$ luminosity is related to the ionizing photon rate as $L_\mathrm{Ly \alpha}^\mathrm{gal} \propto f_\mathrm{Ly \alpha} \left(1-f_\mathrm{esc}\right) \dot{N}_\mathrm{ion}$, where the fraction $f_\mathrm{Ly \alpha}$ of Ly$\alpha$ photons not absorbed by dust is taken as in~\citet{Hayes:2011} as a redshift-dependent parametrisation, while the escape fraction $f_\mathrm{esc}$ follows~\citet{Razoumov:2010} to depend both on redshift and a given halo mass. We populate halos using a parametrised SFR-mass-relation, which is matching the observed trend of an  increasing SFR for smaller mass halos, almost constant for larger halo  masses, while ensuring a reasonable reionization history. For Ly$\alpha$ this yields a luminosity function compatible with observations~\citep{Silva12}. To model the galactic contribution of H$\alpha$ alongside, we make use of the relation between  SFR  and  H$\alpha$ luminosity  from~\citet{Kennicutt:1997ng}.  

Importantly, as Ly$\alpha$ is a resonant line, IGM  attenuation  due  to  its damping  tail needs  to  be  taken  into  account. We  relate  the  intrinsic  luminosity $L_\mathrm{Ly \alpha}^\mathrm{gal}$ assigned  to halos of a given mass as described above to the observed luminosity via IGM optical depth $\tau_\mathrm{Ly\alpha}$ as $L_\mathrm{Ly \alpha, obs}^\mathrm{gal}=L_\mathrm{Ly \alpha}^\mathrm{gal} \exp^{-\tau_\mathrm{Ly\alpha}}$. The IGM optical depth is computed by tracing through the HI density and velocity fields of the 21cm simulation along a chosen line-of-sight (LOS) direction. \rev{When tracing the radiation from an emitting source surrounded by ionised medium this radiation is redshifted between the emission and the edge of neutral medium around, and therefore gets shifted from the line core in resonance to the line wings of lower optical depth on the way to the observer~\citep{Escude98}. We briefly describe the calculation and formulae used to compute $\tau_\mathrm{Ly\alpha}$ in Appendix~\ref{app:damping}; see also~\citet{Heneka17} for a more detailed discussion of the effect of damping on the Ly$\alpha$ power spectrum. Furthermore, the galactic Ly$\alpha$ emission itself is extended due to the spatial and frequency diffusion of Ly$\alpha$ photons when scattering off neutral hydrogen resonantly. These diffuse Ly$\alpha$ halos are an interesting signal to characterise the CGM. As shown in the CDIM Science Report~\citep[chapter 2.4.1]{CDIMreport}, using Ly$\alpha$ radiative transfer calculations, it is possible for a CDIM-type mission to detect a Ly$\alpha$ halo signal and measure its brightness profile towards the end of reionisation when stacking images of galaxies detected in H$\alpha$. We note that we ignore the presence of diffuse Ly$\alpha$ halos for the galactic contribution to brightness in Ly$\alpha$ in this study, while accounting for diffuse and scattered IGM emission as described below.
}

What we call diffuse IGM emission stems from ionising radiation that escapes the halos of ionising sources like line-emitting galaxies and ionizes the neutral medium around. For example Ly$\alpha$ radiation is re-emitted through recombinations after neutral hydrogen has been ionized. The number density of recombinations $\dot{n}_\mathrm{rec}$ in the diffuse IGM is proportional to the comoving free electron density $n_\mathrm{e}$ and the comoving number density of ionized hydrogen $n_\mathrm{HII}$, $\dot{n}_\mathrm{rec} = \alpha n_\mathrm{e} n_\mathrm{HII}$. The recombination coefficient $\alpha$ depends on the kinetic gas temperature. The number density of recombinations is in turn directly proportional to the line luminosity density due to recombinations $l_\mathrm{rec}^\mathrm{IGM}=f_\mathrm{rec}\dot{n}_\mathrm{rec}E_\mathrm{line}$, that we translate to surface brightness for mock observational intensity maps. Diffuse IGM emission due to ionized hydrogen recombining in different line channels is set by the fraction of line photons emitted per hydrogen recombination $f_\mathrm{rec}$ for a line with rest-frame energy $E_\mathrm{line}$. For the Ly$\alpha$-line we set $f_\mathrm{rec}\approx0.66$, and for H$\alpha$ we insert $f_\mathrm{rec}\approx 0.5$, at respective rest-frame energies of $E_\mathrm{Ly\alpha}=1.637 \times 10^{-11}$erg and \rev{$E_\mathrm{H\alpha}=3.028 \times 10^{-12}$}erg (1 erg = $10^{-7}$ Joule). Due to the ratio \rev{between $f_\mathrm{rec}E_\mathrm{line}$ for Ly$\alpha$ to H$\alpha$ of $\sim 7.4$, the diffuse IGM emission of H$\alpha$ is small as compared to Ly$\alpha$ emission in the IGM. We find a mean surface brightness $\nu \mathrm{I}_{\nu}$ of diffuse IGM emission at $z=7$ ($z=10$) in Ly$\alpha$ of $\sim 2.9\times 10^{-9}$ ($2.5\times10^{-11}$) erg$\,$s$^{-1}$cm$^{-2}$sr$^{-1}$ as compared to $\sim 4.1\times 10^{-10}$ ($3.5\times10^{-12}$) erg$\,$s$^{-1}$cm$^{-2}$sr$^{-1}$ in H$\alpha$. For a comparison of the full power spectra we refer to Appendix~\ref{app:Pks}.}

In the case of the resonant Ly$\alpha$ line, also a scattered IGM radiation background exists that extends to the neutral or partially ionized IGM. The main contributors are X-ray excitation of neutral hydrogen and direct stellar emission in the UV between the Ly$\alpha$ frequency and the Lyman-limit, redshifting into Lyman-n resonance. Our simulations take into account hard X-ray sources exciting the IGM as well as emission due to stellar emissivity in the UV, estimated by summing over Lyman resonances.

\begin{figure*}
\centering
\includegraphics[width=0.5\columnwidth,height=0.5\columnwidth,keepaspectratio]{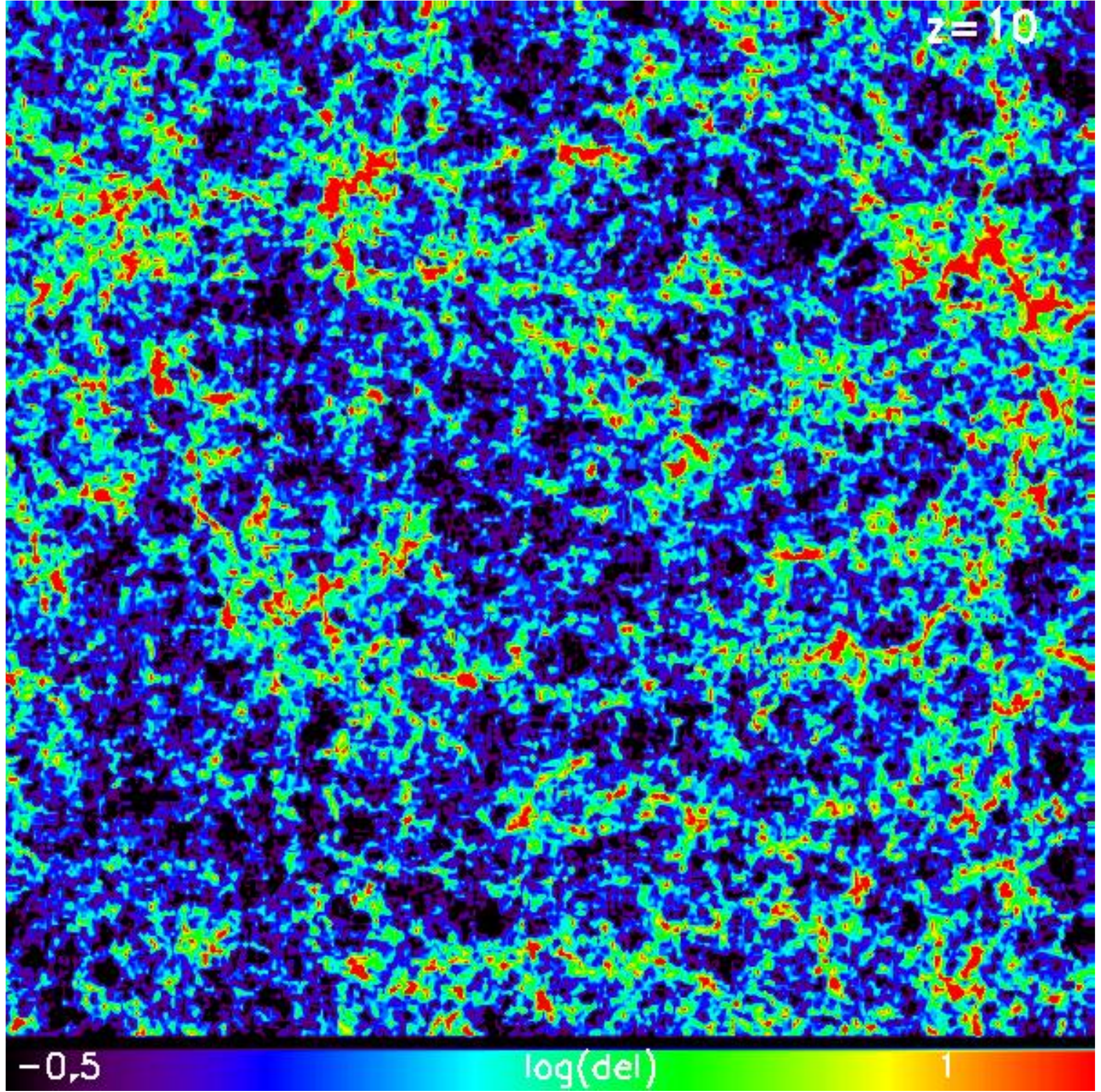}
\hspace{0.01cm}
\includegraphics[width=0.5\columnwidth,height=0.5\columnwidth,keepaspectratio]{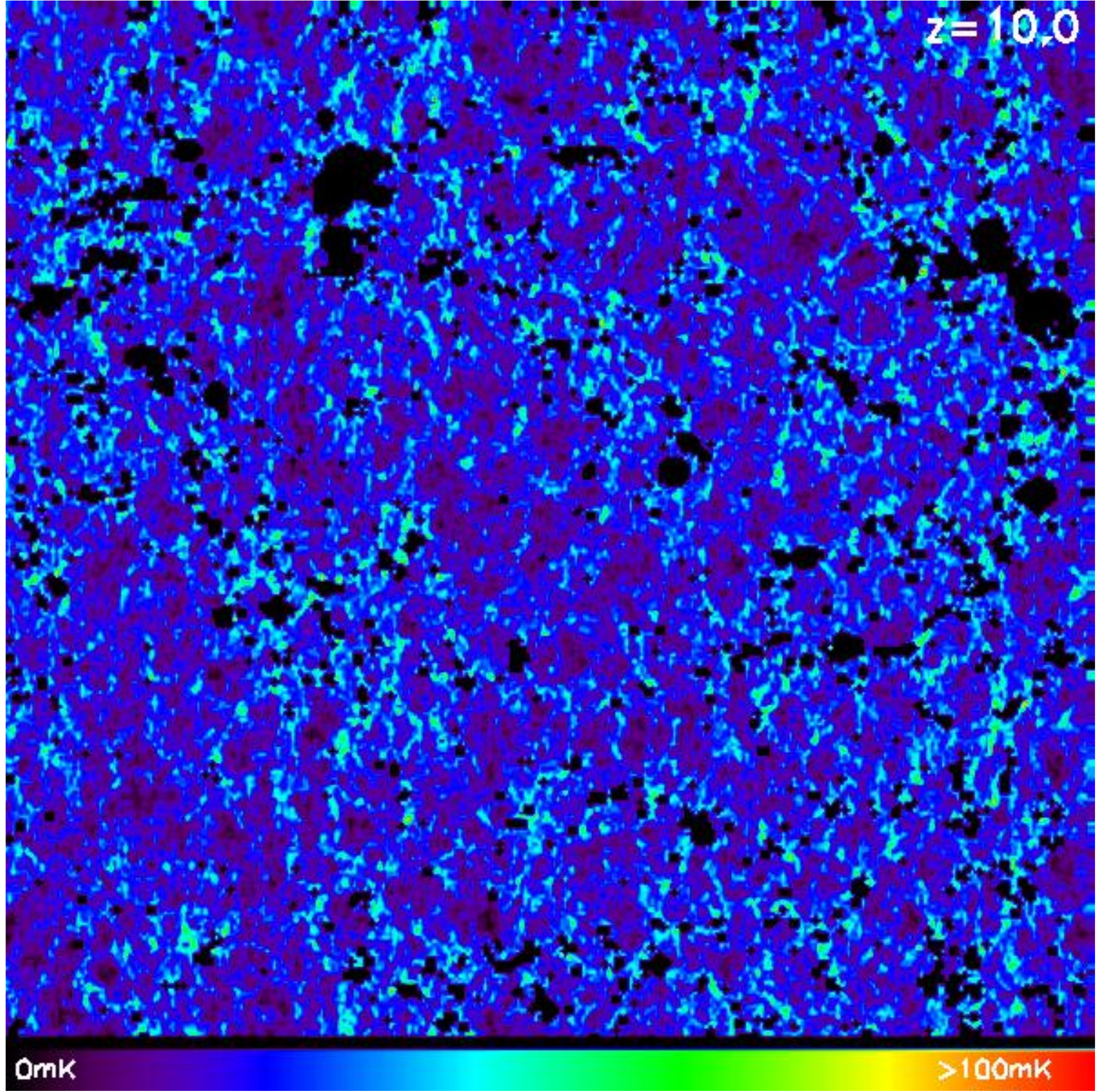}
\hspace{0.01cm}
\includegraphics[width=0.5\columnwidth,height=0.5\columnwidth,keepaspectratio]{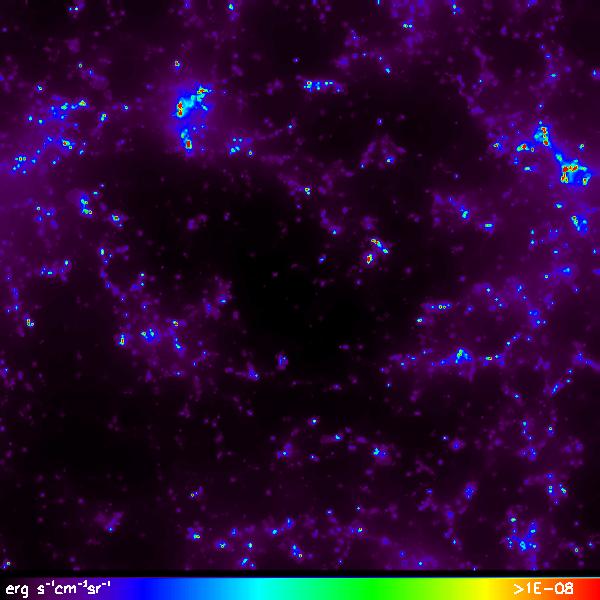}
\hspace{0.01cm}
\includegraphics[width=0.5\columnwidth,height=0.5\columnwidth,keepaspectratio]{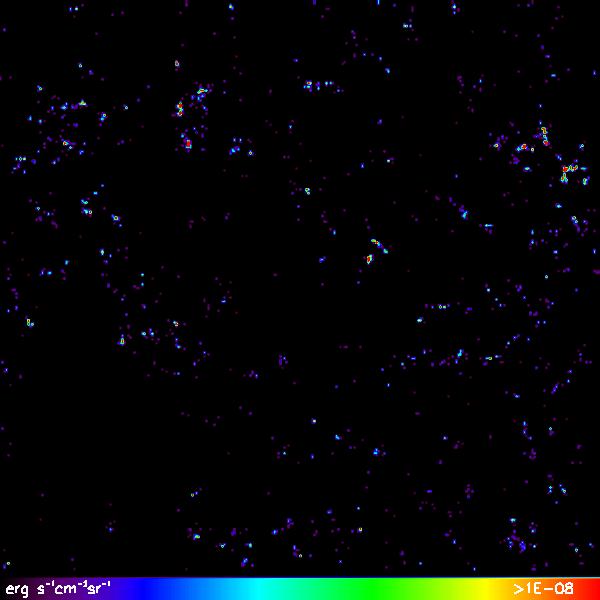}
\vspace{0.01cm}
\vfill
\includegraphics[width=0.5\columnwidth,height=0.5\columnwidth,keepaspectratio]{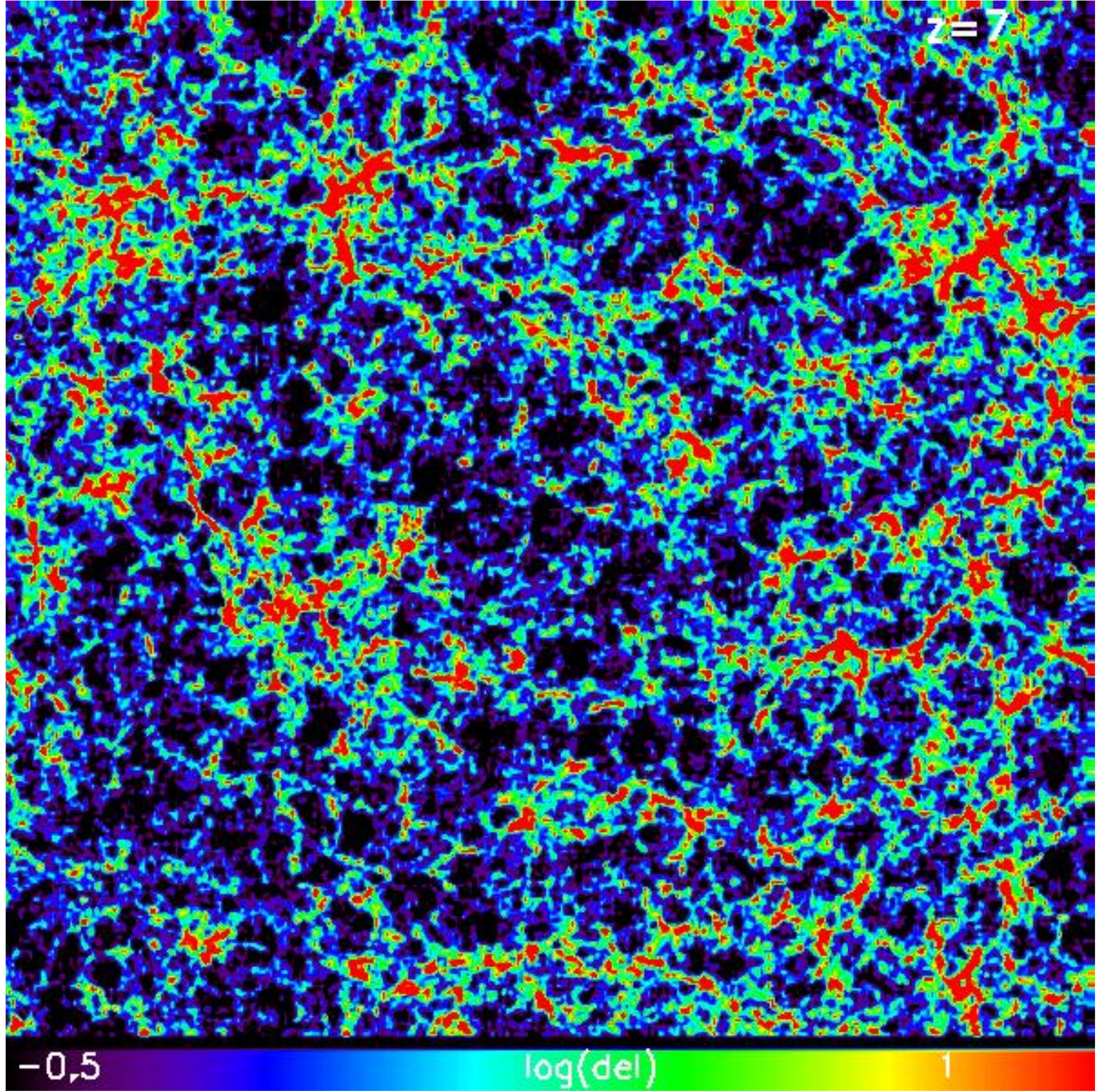}
\vspace{0.01cm}
\includegraphics[width=0.5\columnwidth,height=0.5\columnwidth,keepaspectratio]{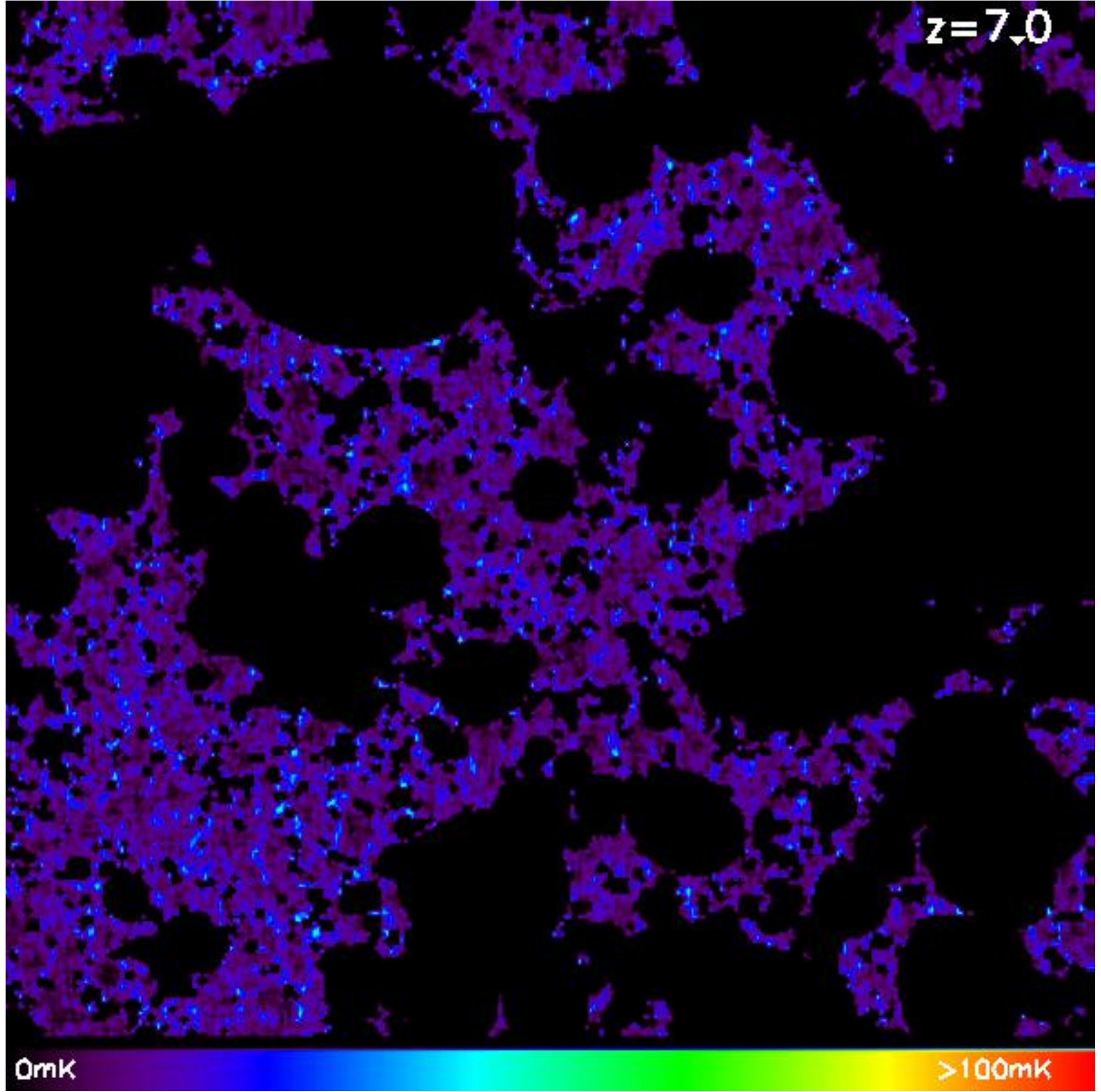}
\hspace{0.01cm}
\includegraphics[width=0.5\columnwidth,height=0.5\columnwidth,keepaspectratio]{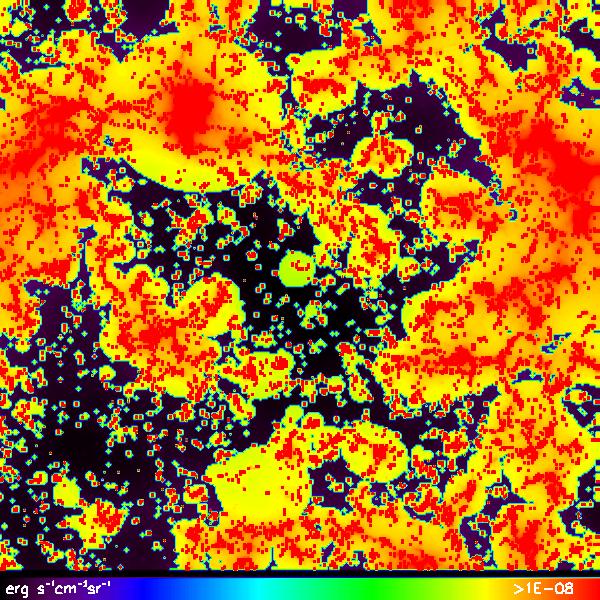}
\hspace{0.01cm}
\includegraphics[width=0.5\columnwidth,height=0.5\columnwidth,keepaspectratio]{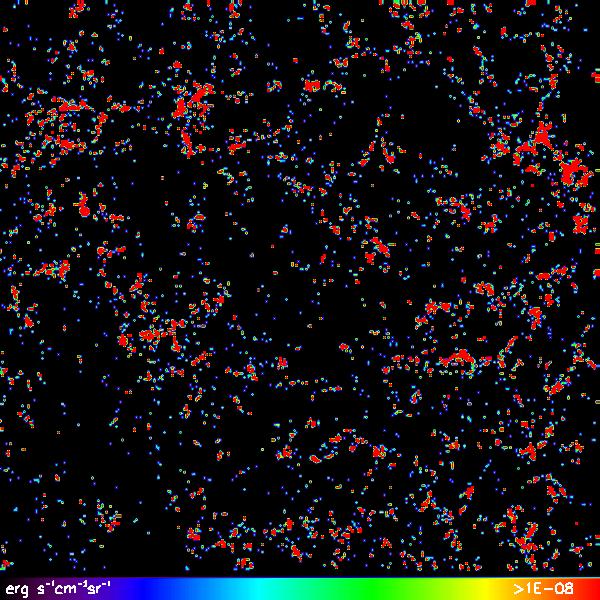}
\caption{Maps of 200$\,$ Mpc box length, from left to right, of the density field, the 21cm brightness temperature, Ly$\alpha$ emission and H$\alpha$ emission, at $z=10$ and $\bar{x}_\mathrm{HI}\sim 0.87$ (top row), $z=7$ and $\bar{x}_\mathrm{HI}\sim 0.27$ (bottom row), for our multi-line IM simulations as described in section~\ref{sec:sim}.}
\label{fig:sim}
\end{figure*}

In Figure~\ref{fig:sim} we show simulation boxes at redshift $z\sim 10$ (top row) and $z\sim 7$ (bottom row), with mean IGM neutral fraction $\bar{x}_\mathrm{HI}\sim 0.87$ and $\bar{x}_\mathrm{HI}\sim 0.27$, respectively, from left to right of the density field, the 21cm brightness offset temperature, the total emission in Ly$\alpha$ and total emission in H$\alpha$. In this model, reionization concludes at around $z\sim 6$. The progress of reionization in 21cm (expansion of 21cm-dark ionised regions) and the complementary increase and expansion of line emission tracing ionized regions as Ly$\alpha$ is clearly visible towards lower redshifts and  approaching the end of reionization.

\section{multi-line IM auto and cross-power}\label{sec:Pk}

After simulation of the line emission of interest and the creation of brightness intensity maps, the fluctuations $\delta_\mathrm{I_\nu}$ in the intensity field $I_\nu \left(\mathbf{x}, z \right)$ at observed frequency $\nu$ are derived for each voxel $\mathbf{x}$ at a given redshift $z$. In Fourier space, the dimensionless, spherically averaged, line power spectrum is then defined as $\tilde{\Delta}^2_\mathrm{I_\nu}\left(k \right)\equiv k^3 / (2\pi^2 V) \langle|\delta_\mathrm{I_\nu}|^2\rangle_\mathrm{k}$ and the corresponding dimensional power spectrum is given by $\Delta^2_\mathrm{I}\left(k \right)= \bar{I}^2_\nu \tilde{\Delta}_\mathrm{I}\left(k \right)$, for volume-averaged intensity $\bar{I}_\nu\left(z\right)$ at redshift $z$. Similarly, between two line fluctuations $\delta_\mathrm{I}$ and $\delta_\mathrm{J}$, we calculate the dimensionless cross-power spectrum as $\tilde{\Delta}^2_\mathrm{I,J}\left(k\right)=k^3 / (2\pi^2 V) \Re \langle \delta_\mathrm{I}\delta^{*}_\mathrm{J}\rangle_\mathrm{k}$, with corresponding dimensional cross-power spectrum $\Delta^2_\mathrm{I,J}\left(k \right)= \bar{I}_\mathrm{I} \bar{I}_\mathrm{J} \tilde{\Delta}_\mathrm{I,J}\left(k \right)$.

To judge the ability to measure both line autopower and cross-power spectra, we calculate their variance based on cosmic variance and the respective multi-line IM instrumental noise power spectrum. For a given mode $\left(\mu, k\right)$ the variance of intensity mapping autopower spectra for line $I$ reads
\begin{equation}\label{eq:autoN}
\sigma^2_\mathrm{I}\left(k,\mu\right) = P_\mathrm{I} \left(k,\mu\right) + P_\mathrm{N, I} \left(k,\mu\right) , 
\end{equation}
with sample variance $P_\mathrm{I}=\langle|\delta_\mathrm{I}|^2\rangle$ and instrument-specific noise power spectrum $P_\mathrm{N, I}$. The variance estimate of  the  intensity mapping cross-power  spectrum for line $I$ and $J$ reads (e.g.~\citet{2007ApJ...660.1030F,2008LidzNoise})
\begin{equation}\label{eq:crossN}
\sigma^2_\mathrm{I,J}\left(k,\mu\right) = \frac{1}{2}\left[ P_\mathrm{I,J} \left(k,\mu\right) + \sigma_\mathrm{I} \left(k,\mu\right) \sigma_\mathrm{J} \left(k,\mu\right) \right] .
\end{equation}
The total variance $\sigma(k)$ for the full spherically averaged autopower and cross-power spectrum is the binned sum of the variance $\sigma_\mathrm{I}(k,\mu)$ and $\sigma_\mathrm{I,J}(k,\mu)$, respectively, over all angles $\mu$ (all modes $k^2 = k^2_\parallel + k^2_\perp$), divided by the respective number of modes per bin. For the calculation of the noise power spectrum we explicitly count the number of modes per bin. 

For a multi-line IM experiment, the noise power spectrum $P_\mathrm{N, I}$ is modelled as a thermal white-noise spectrum and a window function $W_\mathrm{I}\left(k,\mu \right)$ that accounts for limited spatial and spectral instrumental resolution. The noise power spectrum reads 
\begin{equation}
P_\mathrm{N, I}=\sigma^2_\mathrm{N}V_\mathrm{vox}W_\mathrm{I}\left(k,\mu \right) , 
\label{eq:P_N}
\end{equation}
where $\sigma_\mathrm{N}$ is the instrument-specific thermal noise variance and $V_\mathrm{vox}$ the comoving voxel size (see section 4, in particular equations (24) and (31) of~\citet{Heneka17} for a more detailed description of 21cm and Ly$\alpha$ instrumental noise power spectra). For parallel modes $k_\parallel = \mu k$ along the LOS and transverse modes $k_\perp = (1-\mu^2 )^{1/2} k$ the window function reads~\citep{2011ApJ...741...70L}
\begin{equation}
   W_\mathrm{I}\left(k,\mu \right) = \mathrm{e}^{\left(k_\parallel / k_\mathrm{\parallel , res}\right)^2 + \left(k_\perp / k_\mathrm{\perp , res}\right)^2}  . 
\end{equation}
The ability of an instrument to resolve a certain region in mode-space (and thus measure fluctuations) crucially depends on the spectral and spatial instrumental resolution in parallel and transverse modes, $k_\mathrm{\parallel , res}$ and $k_\mathrm{\perp , res}$. For survey sensitivity studies it is thus important to note that
\begin{align}\label{eq:noisep}
k_\mathrm{\parallel , res} &\propto R \\ \nonumber
k_\mathrm{\perp , res} &\propto  \frac{1}{x_\mathrm{pix}}  \left( \propto \frac{1}{\theta_\mathrm{min}} \right), 
\end{align}
where $R$ is the spectral resolution and $x_\mathrm{pix}$ ($\theta_\mathrm{min}$) are the spatial pixel size (angular beam size). Note that both $x_\mathrm{pix}$ and $R$ enter the multi-line IM noise power spectrum as well via $V_\mathrm{vox}=A_\mathrm{pix}r_\mathrm{pix}$ with pixel area $A_\mathrm{pix}$ and comoving pixel depth $r_\mathrm{pix}=\chi(R)$. In section~\ref{sec:survey} we will showcase how the detectability of autopower and cross-power intensity mapping spectra depends on the survey-specific sensitivity given by the thermal noise variance $\sigma_\mathrm{N}$, the spectral resolution $R$ and the spatial pixel size $x_\mathrm{pix}$ for the example of Ly$\alpha$ and 21cm intensity mapping.

\begin{figure*}
{\setlength{\tabcolsep}{0.05cm}
\begin{tabular}{ccc}

\includegraphics[width=0.65\columnwidth]{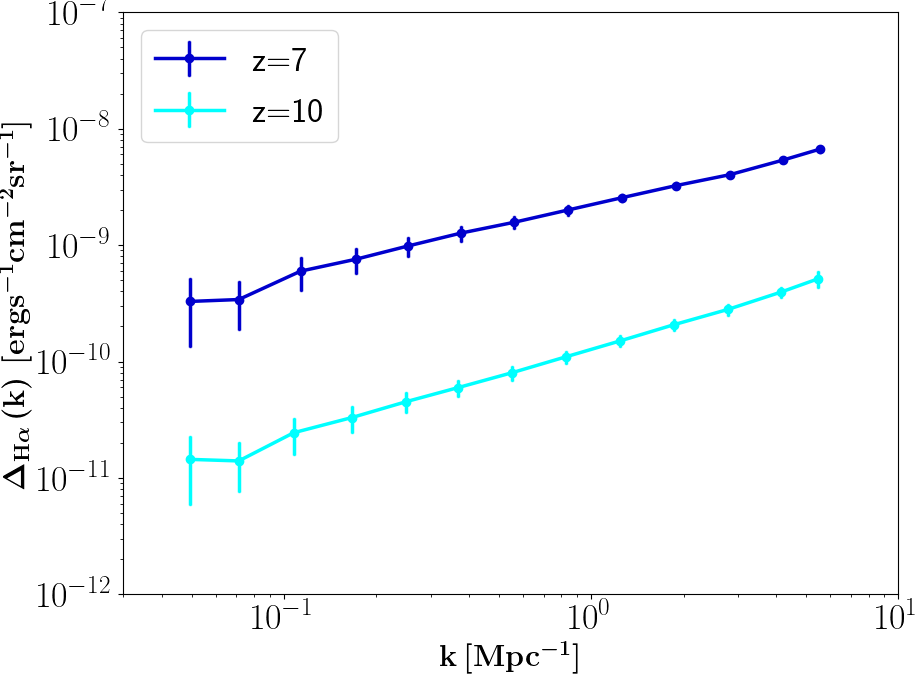} \put(1,100){H$\alpha$} \hfill&\hfill& 
 \\
\includegraphics[width=0.65\columnwidth]{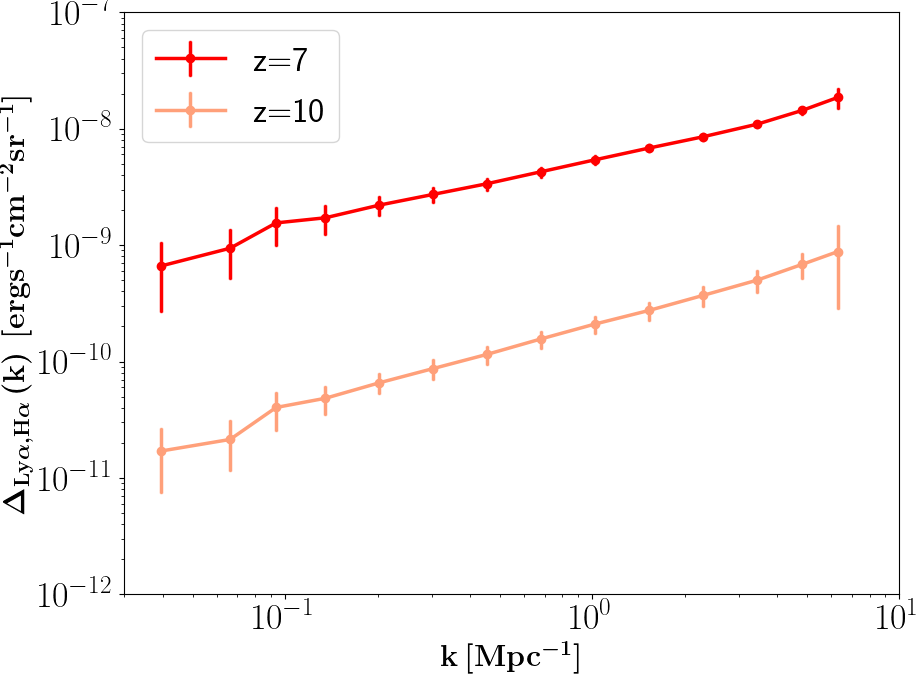} \hfill &
\includegraphics[width=0.65\columnwidth]{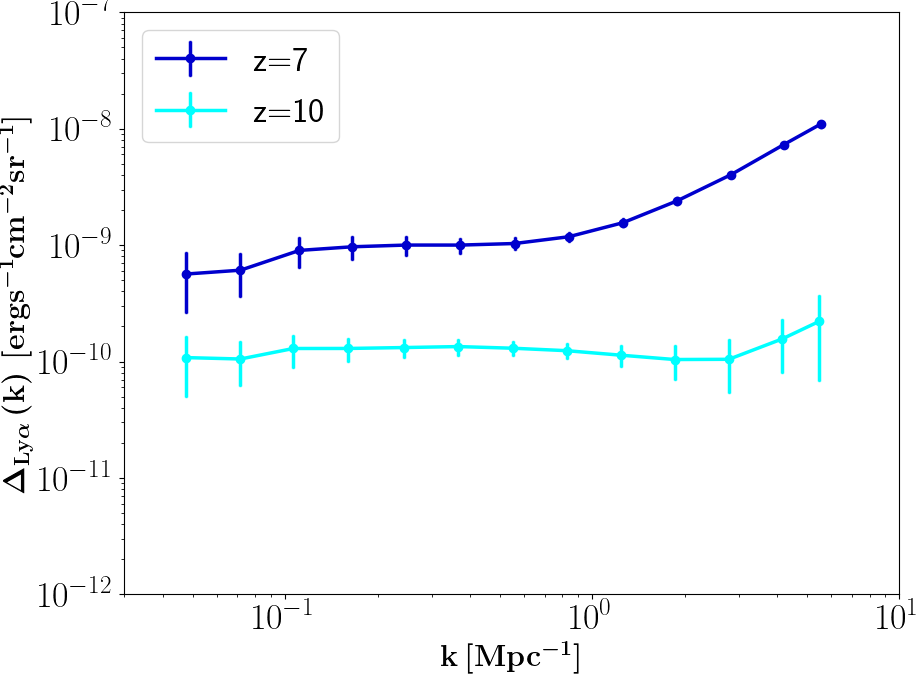}  \put(1,100){Ly$\alpha$} \hfill &  \\
\includegraphics[width=0.65\columnwidth]{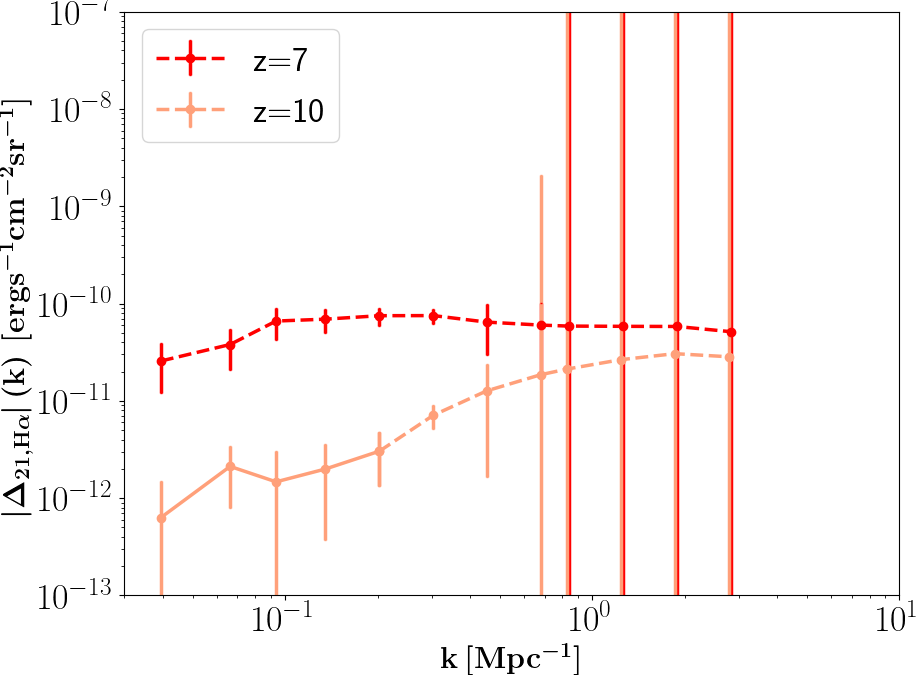} \hfill&
\includegraphics[width=0.65\columnwidth]{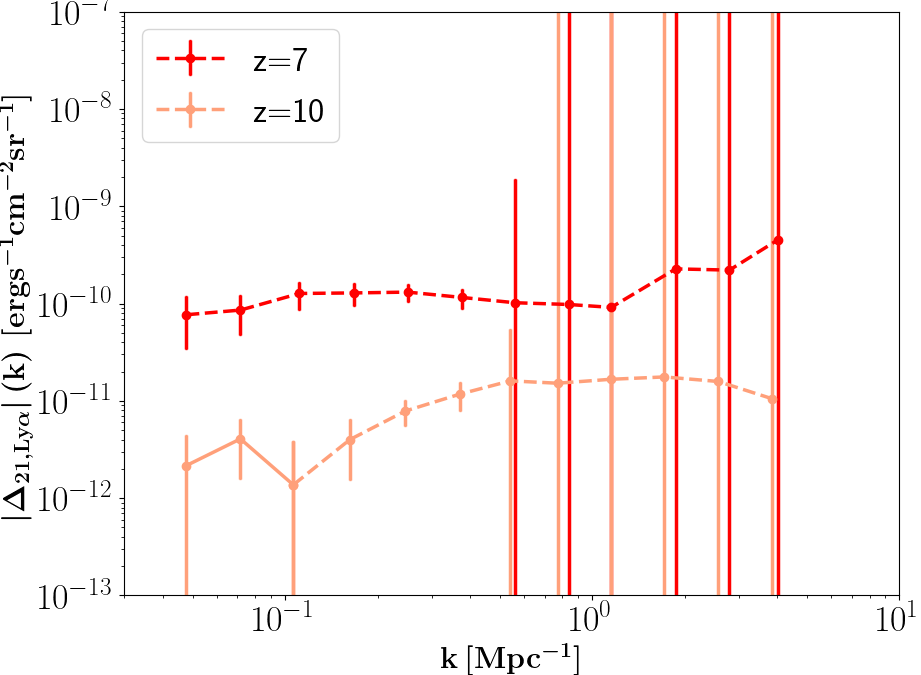}  \hfill&
\includegraphics[width=0.65\columnwidth]{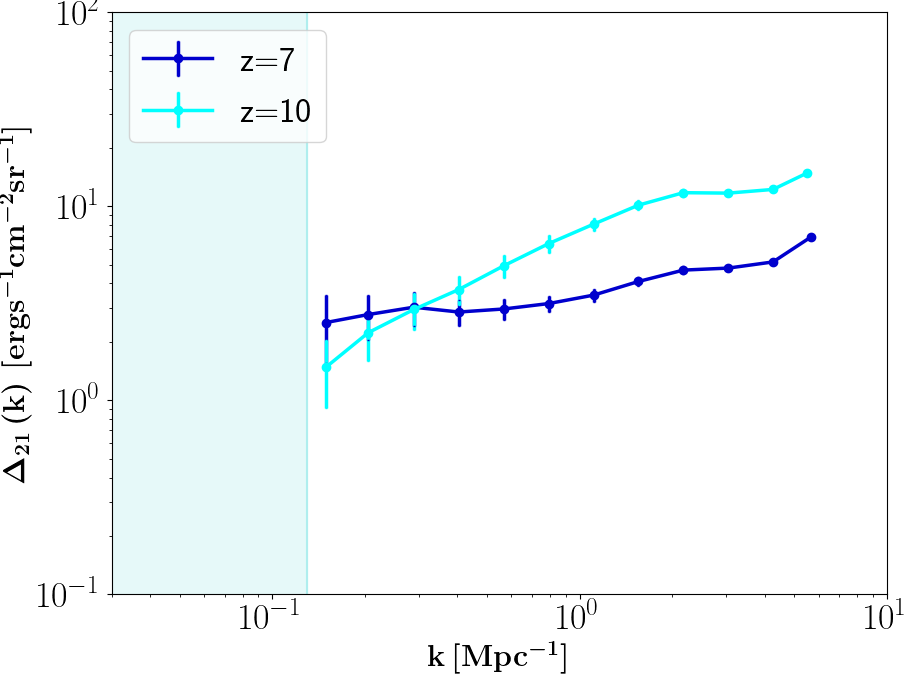} \put(1,100){21cm}
\end{tabular}
}
\caption{Autopower (top panels of each column, blue at $z=7$, cyan at $z=10$) and cross-power spectra (red at $z=7$, orange at $z=10$) for 21cm, Ly$\alpha$ and H$\alpha$ lines; error-bars are shown at 1$\sigma$ level for the CDIM medium survey and SKA1-Low instrument specifics (as described in section~\ref{sec:Pk}). The dashed lines for 21cm x Ly$\alpha$ and 21cm x H$\alpha$ panels indicate negative correlation. The light blue shaded region for 21cm is lost due to foregrounds, as are modes contained in the so-called foreground wedge.}
\label{fig:tri}
\end{figure*}

Our fiducial mission for Ly$\alpha$ and H$\alpha$ intensity mapping follows the expected design of the Cosmic Dawn Intensity Mapper \rev{as presented in the CDIM Science Report}~\citep[CDIM]{CDIMreport}. We assume a frequency resolution of $R=300$, a current best estimate for the instrumental line sensitivity \rev{for the planned medium-sized survey} of $\sigma_\mathrm{N}=1.48\times 10^{-19}$erg s$^{-1}$cm$^{-2}$Hz$^{-1}$sr$^{-1}$ ($1.26\times 10^{-19}$erg s$^{-1}$cm$^{-2}$Hz$^{-1}$sr$^{-1}$) for Ly$\alpha$, and $8.0\times 10^{-19}$erg s$^{-1}$cm$^{-2}$Hz$^{-1}$sr$^{-1}$ ($1.1\times 10^{-18}$erg s$^{-1}$cm$^{-2}$Hz$^{-1}$sr$^{-1}$) for H$\alpha$, at redshift $z=7$ (redshift $z=10$) for mirror size $D_\mathrm{ap}=0.83\,$m and pixel size of $x_\mathrm{pix}=1''$, as well as a survey area of $\sim 31$ deg$^2$ for the CDIM medium-size survey over the wavelength range 0.8--9 microns. In section~\ref{sec:survey} we also compare the CDIM-type setup to an all-sky SPHEREx-type survey, with an aperture of 20 cm, pixel size of $x_\mathrm{pix}=6.2''$ and with wavelength range 0.75--5 microns at spectral resolution $R=41$. Foreground residuals due to lower-redshift interloping line emission, for example in the case of Ly$\alpha$ interloping H$\alpha$, OII, OIII lines, can be projected down to levels far below (factor 10--100) the expected multi-line IM signals by  low-level mitigation strategies, such as high flux masking (see~\citet{2017ApJ...835..273G,2019BAAS...51g..23Cf}). 

\rev{As the target sensitivity $\sigma_\mathrm{N}$ is directly driven by the respective science requirements and is itself dependent on a range of instrumental and survey parameters, we use it as an effective parameter in this study. We nevertheless would like to discuss some main dependencies here. Important parameters are the mirror or aperture size $D_\mathrm{ap}$ and the pitch $p_\mathrm{pix}$ of a detector pixel that determine the angle subtended by detector elements, or spatial pixel size to be $\propto p_\mathrm{pix}/D_\mathrm{ap}$. Besides spatial resolution attainable, this affects for example the photocurrent from the sky background. Here larger aperture (pitch) results in smaller (larger) flux uncertainty. The total photocurrent that enters the rms noise of the detector consists besides dark current of the telescope background $i_\mathrm{instr}$ and the sky background $i_\mathrm{sky}$ terms. For example at low temperatures the CDIM detector is operating at ($\sim 35\,$K), the background is dominated by scattered solar and thermal zodiacal light, that can be modelled as black body spectra. In the case of CDIM this is mitigated with a high number of redundant visits of the same fields. The actual photocurrent is then proportional to these background levels modulo subtended angles and optical, detector and filter efficiencies. Lastly, besides with total photocurrent and sky background, the rms detector noise is inversely proportional to the integration time on-sky $t_\mathrm{int}$ and scales with detector read noise $\mathrm{dQ}$ (which in turn depends on detector sampling and integration times via $\mathrm{dQ}\propto \sqrt{T_\mathrm{samp}/t_\mathrm{int}}$). \footnote{For a base public version of a sensitivity calculator see for example https://github.com/zemcov/CDIM$\_$APROBE}}

\rev{As fiducial} 21cm observations we assume \rev{an SKA1-Low instrument setup} with 1000h on-sky integration, for 6 hours per night tracked scan during 167 days per year. The thermal  noise  depends  on  the  characteristics  of  the radio  interferometer, where we use SKA1-Low characteristics as detailed in the SKA1 \textit{System Baseline Design} document.\footnote{https://astronomers.skatelescope.org/wp-content/uploads/2016/05/SKA-TEL-SKO-0000002\_03\_SKA1SystemBaselineDesignV2.pdf} The sky temperature is assumed to scale as $T_\mathrm{sky}=60\lambda^{2.55}$mK with wavelength $\lambda$, and the system temperature to follow the relation $T_\mathrm{rec}=1.1 T_\mathrm{sky}+40$mK. \rev{For comparison in section~\ref{sec:survey} we briefly explore a fiducial HERA instrumental setup with antennae distributed on a hexagonal grid and 1000h integration in drift mode~\citep{DeBoer:2016tnn}.}

In our default foreground treatment the modes in \rev{an extended (by 0.1 hMpc$^{-1}$ in $k_{\parallel}$}) foreground wedge are discarded, which corresponds to the moderate 'mod' foreground option in the publicly available 21cmSense code~\citep{Pober:2013,Pober:2013jna}.\footnote{https://github.com/jpober/21cmSense} 
In section~\ref{sec:survey} we test for comparison the optimistic 'opt' foreground option provided in 21cmSense, where only modes inside the primary field of view \rev{(of the 21cm experiment)} are excluded to thus enlarge the detectable EoR window in 21cm. \rev{Also, a horizontal portion of the foreground wedge due to spectrally smooth foregrounds is predicted to swamp modes at low $k_\mathrm{\parallel}$ (e.g. below $k_{\parallel}\sim 0.06\,$Mpc$^{-1}$, see~\citet{Dillon14}). Note though, that our two foreground scenarios, 'mod' and 'opt' bracket the impact of such a cut in low $k_\mathrm{\parallel}$ modes, as in the moderate scenario the entire wedge is shifted upward in $k_\mathrm{\parallel}$ such that for example at $k_{\perp}\sim 0.01\,$Mpc$^{-1}$ automatically $k_\mathrm{\parallel}$ modes below $\sim 0.1\,$Mpc$^{-1}$ are discarded, while the 'opt' scenario includes a relatively low cut in $k_\mathrm{\parallel}$ for our k-range of interest.}

\begin{figure*}
\centering
\includegraphics[width=0.32\textwidth]{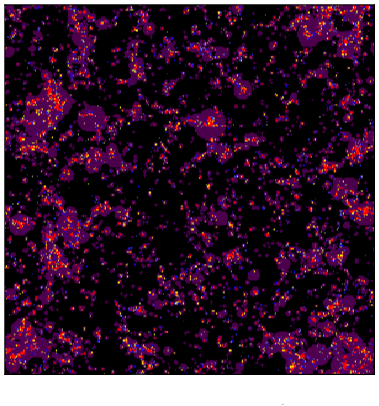}
\includegraphics[width=0.32\textwidth]{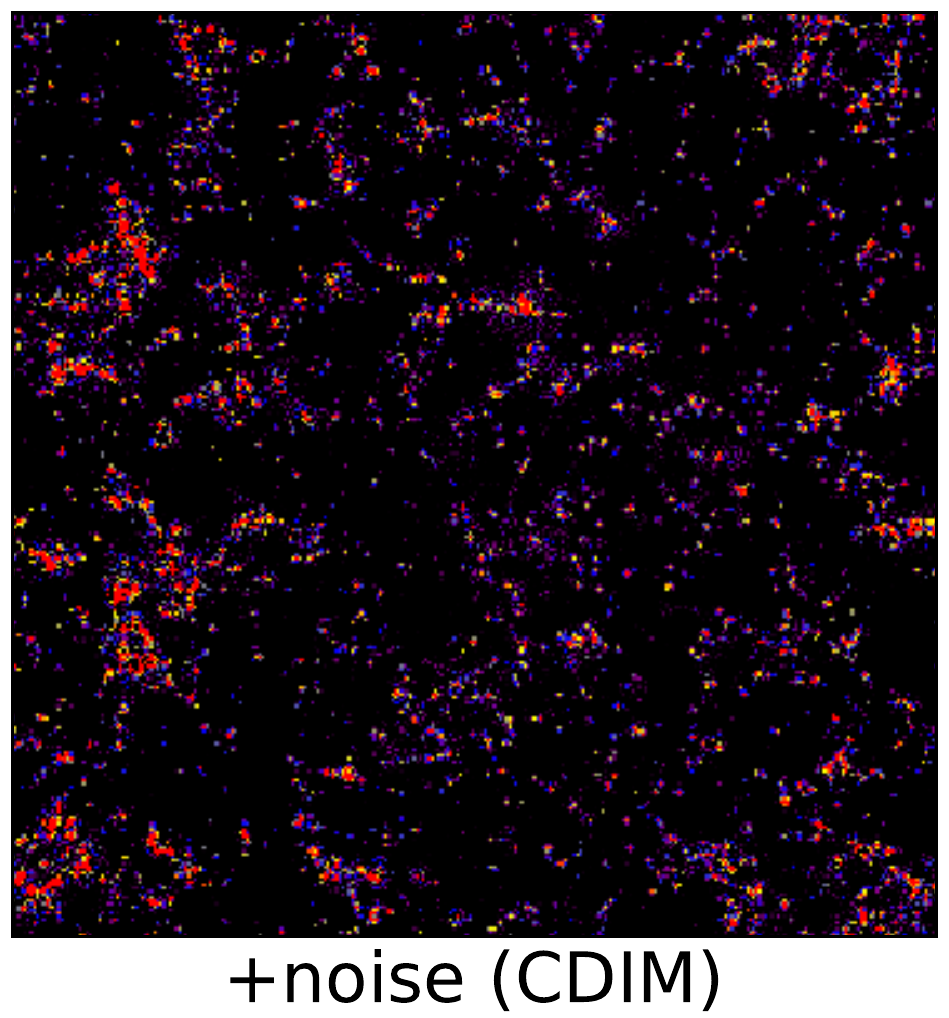}
\includegraphics[width=0.32\textwidth]{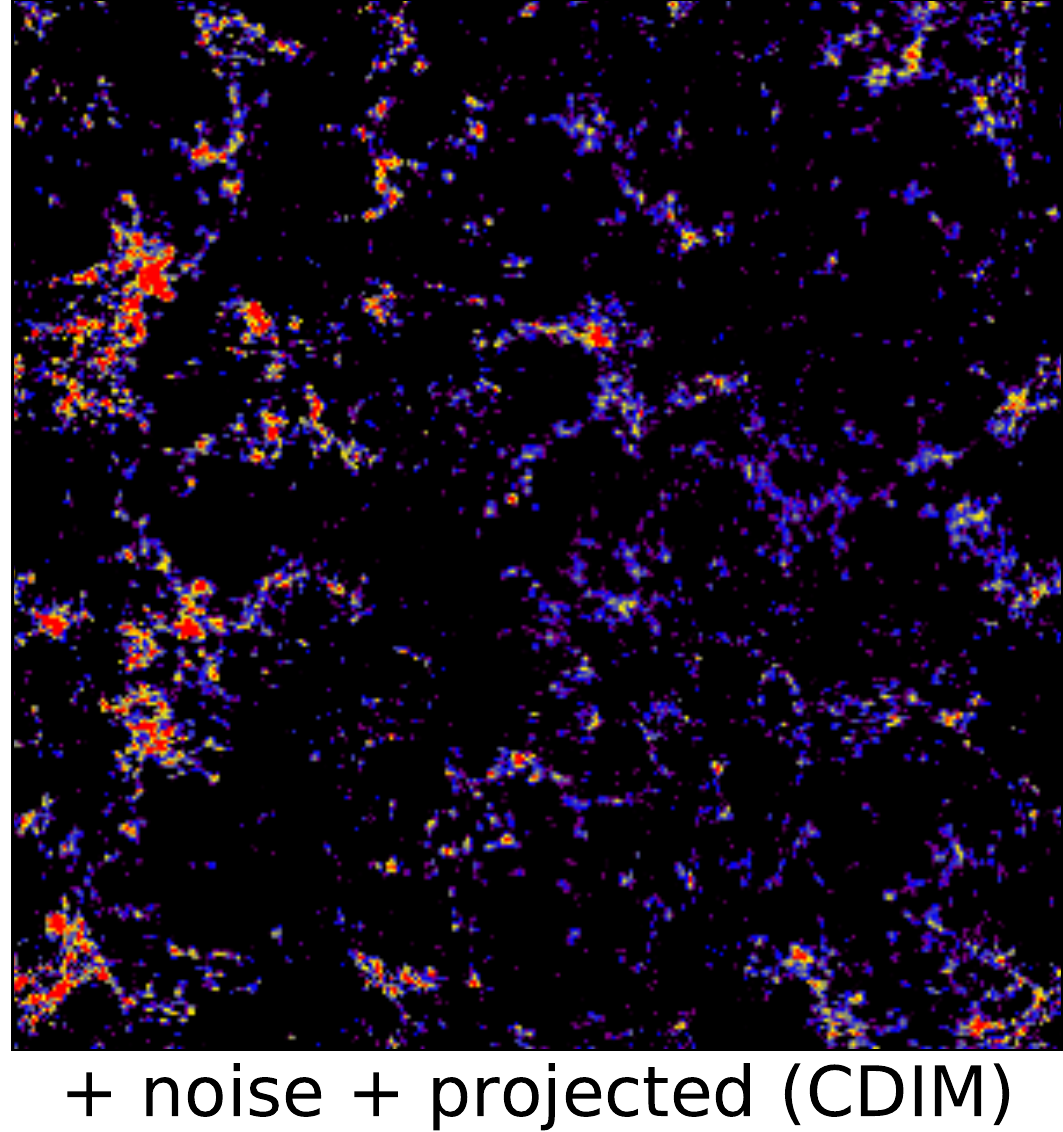}
\caption{Left: Predicted Ly$\alpha$ signal at $z\sim8$ around the midpoint of reionization with mean IGM neutral fraction $\mathrm{\bar{x}}_\mathrm{HI}\sim 0.5$; Middle: including the thermal noise model as well as instrumental noise due to limited spatial and spectral resolution for a CDIM-like mission, but per simulation slice in frequency; Right: projection including thermal and instrumental noise that matches the frequency resolution of a CDIM-like mission (spectral resolution R=300 meaning a slicing of $\sim 15\,$Mpc width at $z=8$). \rev{The signal neglects foreground contamination in the Ly$\alpha$ survey.}}
\label{fig:mock}
\end{figure*}

Figure~\ref{fig:tri} showcases as a triangular plot 21cm, Ly$\alpha$, and H$\alpha$ autopower (blue and cyan) and cross-power spectra (red and orange) derived from our simulations (described in section~\ref{sec:sim}). The error-bars are plotted at the 1$\sigma$ level and follow Eq.~(\ref{eq:autoN}) and Eq.~(\ref{eq:crossN}), to include cosmic variance as well as thermal and instrumental noise. We plot the projected power spectra, where the modes 'lost' due to limited spectral and spatial resolution are projected out. Besides the high signal-to-noise measurement of the 21cm autopower spectrum during reionization in reach for the SKA, a multi-line intensity mapping mission similar to CDIM is able to provide us with complementary high signal-to-noise measurements of emission line autopower spectra and cross-power spectra, in this case for Ly$\alpha$ and H$\alpha$ lines up to the early stages of reionization. Both the cross-correlation of Ly$\alpha$ and H$\alpha$ emission with the 21cm signal is detectable over more than one order in scale $k$ up to redshifts of $z\sim 10$. The characteristic negative correlation and shape change from late stages of reionization towards its midpoint, tracing the expansion of ionized regions, reaching positive correlation at large scales during early stages of reionization, turning over to negative at smaller, typically already ionized, scales. The possible high signal-to-noise measurement, picking up the total line emission fluctuations present for a line, as well as the complementarity to the 21cm measurement and characteristic evolution of the cross-signal during reionization makes multi-line IM a valuable tool to increase information gain on emission sources and strength during the EoR, while serving as an important cross-check for the cosmological nature of the 21cm signal.

\section{Spectral line IM mock observations}\label{sec:mocks}

For the creation of emission line IM mock observations displayed here, we assume a CDIM-type IM mission. The associated noise is calculated for instrument specifics as described in the previous section. The noise estimate includes thermal as well as instrumental noise following Eq.~(\ref{eq:P_N}). We  generate the mock IM maps by randomly sampling the above mentioned noise power in Fourier space, and adding it to the simulated signal.  Specifically for Ly$\alpha$ IM we show in Figure~\ref{fig:mock} the simulated signal at $z\sim8$ at the midpoint of reionization (left panel), including the noise model for CDIM (middle panel) and projecting the mock observation for a redshift binning as expected for CDIM-like spectral resolution (right panel). Note that the field of view extends beyond the depicted box length of 200$\,$ Mpc. \rev{Also, foreground contamination is neglected here; we expect though foreground residuals due to lower-redshift interloping line emission to be projectable far below the expected Ly$\alpha$ signal.} As can be clearly seen in the right panel, such a multi-line IM mission is able to map the large-scale structure (LSS) in Ly$\alpha$ emission both picking up the more confined galactic emission as well as more diffuse emission, especially for lower density regions of the LSS. \rev{In Appendix~\ref{app:Pks} we show the theoretical power spectra of the galactic and IGM contributions to Ly$\alpha$ emission for reference.}

\begin{figure*}
\centering
\includegraphics[width=1.0\columnwidth]{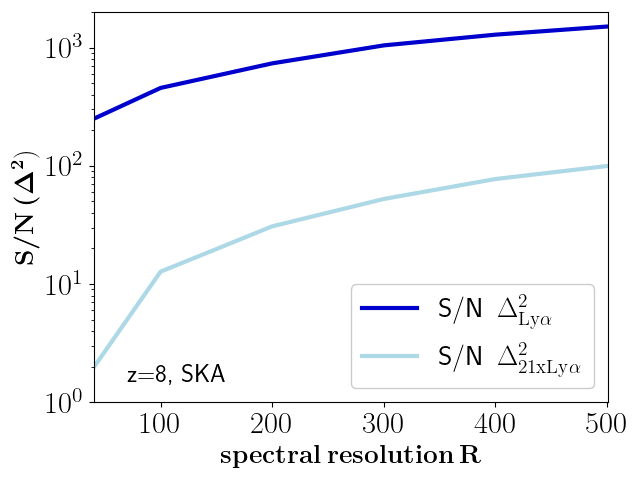}
\includegraphics[width=1.0\columnwidth]{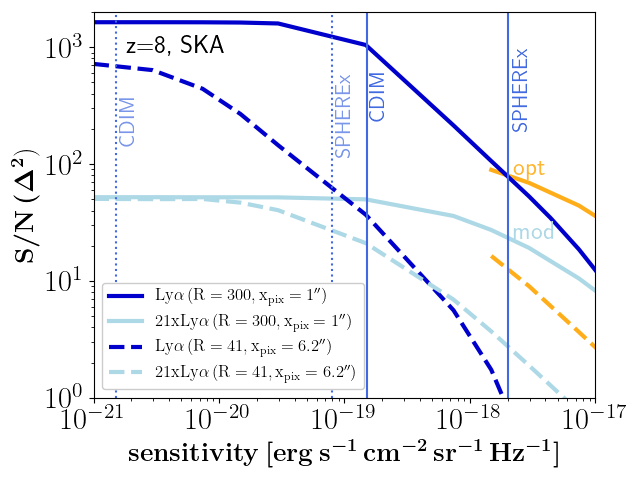}
\includegraphics[width=1.0\columnwidth]{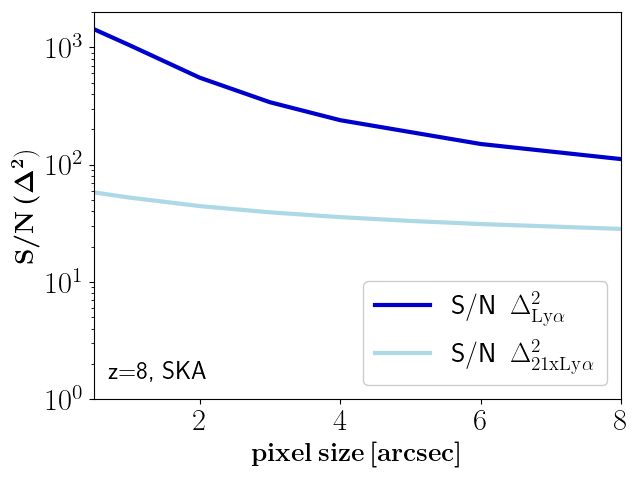}
\includegraphics[width=1.0\columnwidth]{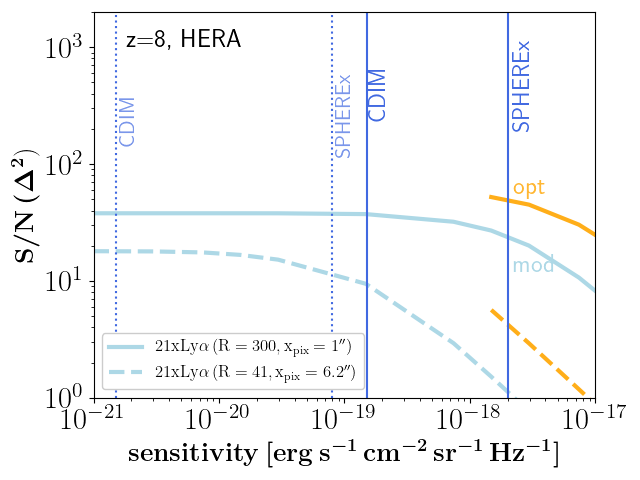}
\caption{Cumulative S/N at $z\sim8$ during the EoR for 21cm-Ly$\alpha$ cross-spectra (cyan) and Ly$\alpha$ auto power (blue) as a function of spectral resolution (top left), line sensitivity (right panels), and pixel size (bottom left). Drawn lines leave the respective non-varied survey parameters fixed to a CDIM-type medium survey set-up (in particular $R=300$ and $x_\mathrm{pix}=1''$), while the dashed lines in the right panels instead assume SPHEREx-type spectral resolution and pixel size ($R=41$ and $x_\mathrm{pix}=6.2''$). For the cross spectra in cyan we assumed moderate 'mod' foreground treatment of the 21cm signal to discard \rev{an extended} foreground wedge region, except for the \rev{orange} broken off lines in the right panels that illustrate the effect of assuming a more optimistic 'opt' 21cm foreground treatment where only modes within the primary field of view of the instrument, SKA1-Low \rev{or HERA (bottom right)}, are discarded, \rev{see~\citet{Pober:2013jna}}. For guidance, the vertical blue lines display the current estimate for the \rev{surface brightness} sensitivity of CDIM at $\sigma_\mathrm{N}=1.5\times 10^{-19}$ and \rev{a minimum requirement of} $\sigma_\mathrm{N} \sim\times 10^{-18}$ erg s$^{-1}$cm$^{-2}$sr$^{-1}$Hz$^{-1}$]$^2$ for SPHEREx. \rev{The respective dotted lines represent current estimates for both surveys that optimise sensitivity.}}
\label{fig:SN}
\end{figure*}

\begin{figure}
\centering
\includegraphics[width=1.03\columnwidth]{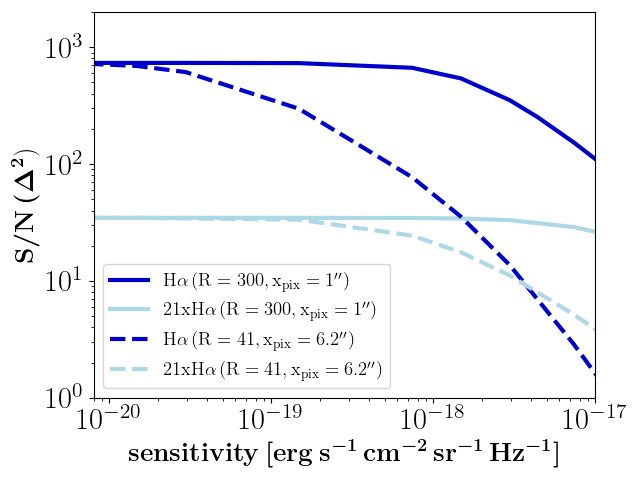}
\caption{Cumulative signal-to-noise at $z\sim8$ during the EoR for 21cm-H$\alpha$ cross spectra with SKA1-Low (cyan) and H$\alpha$ auto power spectra (blue) as a function of line sensitivity. Drawn lines leave the respective non-varied survey parameters fixed to a CDIM-type medium survey size set-up (in particular $R=300$ and $x_\mathrm{pix}=1''$), while the dashed lines assume spectral resolution and pixel size as for SPHEREx. Note though, that as SPHEREx covers wavelengths up to 5 microns it does not have the range for H$\alpha$ at $z>5$. For the cross spectra in cyan we assume 'moderate' foreground treatment of the 21cm signal to discard the foreground wedge region. Note as well, that for H$\alpha$ the dependency of the cumulative S/N on spectral resolution and pixel size is less pronounced than for Ly$\alpha$ at roughly the same order of magnitude in S/N.}
\label{fig:SNHa}
\end{figure}

\begin{figure}
\centering
\includegraphics[width=\columnwidth]{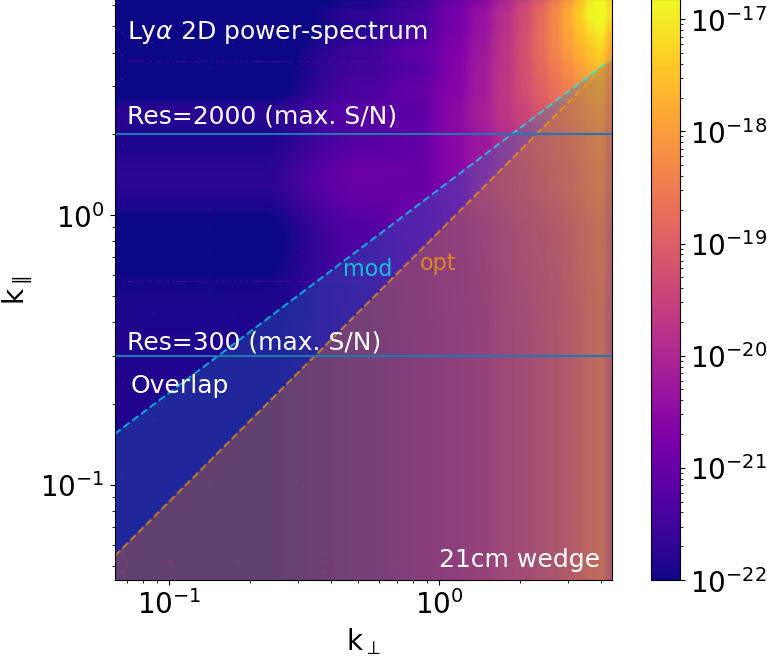}
\caption{Ly$\alpha$ 2D power spectrum $\Delta^2\left(k_\perp, k_\parallel \right)$ at $z=8$, color-coded in units of [erg s$^{-1}$cm$^{-2}$sr$^{-1}$Hz$^{-1}$]$^2$. The shaded area (lower right) is the 21cm wedge; \rev{the shaded area below the orange dashed line labeled 'opt' corresponds to a optimistic 21cm foreground scenario where the wedge is restricted to the primary beam FWHM, whereas the shaded area below the blue dashed line labeled 'mod' corresponds to a moderate 21cm foreground scenario where the wedge is more extended.} Horizontal lines correspond to resolution limits that restrict measurements to the area below; note that the horizontal lines is chosen to maximise the S/N noise of the cross-signal, leading to slight deviations from the native resolution translated into $k_\parallel$.'Overlap' shows exemplary where the region accessible jointly for 21cm and Ly$\alpha$ measurements is located, left of the wedge region and below the horizontal resolution line.}
\label{fig:SN-wedge}
\end{figure}

\section{Exploration of survey parameters}\label{sec:survey}
Large-scale three-dimensional (tomographic) intensity maps of reionization and the cosmic dawn as traced by emission lines are key to constrain the process of reionization and properties of sources that dominate the ionising photon budget. This can be translated to the ability to provide high signal-to-noise measurements of line fluctuation power spectra in fine redshift intervals. In the following we investigate signal-to-noise ratios attainable for both emission line autopower, and cross-power spectra with 21cm, measurable by a wide-field intensity mapping mission. We vary critical instrument specifics in order to identify optimal design routes for synergies of multi-line IM with 21cm IM.

\subsubsection*{\rev{Parameter dependency of cumulative S/N}}

From our multi-line IM simulations as described in section~\ref{sec:sim}  autopower and cross-power spectra are extracted and noise is estimated as outlined in the previous sections. We vary instrument line sensitivity $\sigma_\mathrm{N}$, frequency resolution $R$ and detector pixel size $x_\mathrm{pix}$ as the parameters our multi-line IM noise power spectrum estimate, see Eqs.~(\ref{eq:P_N}) and ~(\ref{eq:noisep}), is sensitive to.  \rev{In addition, for a discussion of best choice for minimising the error bars on the power spectra of interest, we compare in Appendix~\ref{app:errors} the sample variance and the noise power spectrum for our two fiducial mission types, CDIM and SPHEREx, that we perturb around. }

In Figure~\ref{fig:SN} we explore at $z\sim 8$ (roughly the midpoint of reionization in our model) the signal-to-noise ratio (S/N) of both the Ly$\alpha$ auto spectrum (blue upper lines) and the 21cm-Ly$\alpha$ cross-power spectrum (cyan lower lines) as a function of spectral resolution $R$ (top left), line sensitivity $\sigma_\mathrm{N}$ (right panels), and pixel size $x_\mathrm{pix}$ (bottom left). Depending on spectral resolution and the corresponding window function, we chose a cut in parallel modes $k_\parallel$ that maximises the cumulative S/N attainable (in our case this means a cut at slightly larger $k_\parallel$ than suggested by $k_\mathrm{\parallel, res}$).

Here, it can be seen, that for the optimal experiment aiming at high-redshift cross-power spectra a spectral resolution of $R>100$ is advantageous for a cumulative S/N>10 per redshift bin (see the cyan line in the top left panel). Furthermore, the cumulative S/N for the Ly$\alpha$ autopower sees a rather steep increase from S/N $\sim 10$ up to S/N $>10^3$ per tomographic redshift bin for line sensitivities up to $\sigma_\mathrm{N} \sim 10^{-19}$erg s$^{-1}$cm$^{-2}$sr$^{-1}$Hz$^{-1}$, while the S/N for the cross-power spectrum with 21cm (SKA1-Low) rises more moderately from S/N $\sim$ 10 to 100 (see both blue and cyan drawn lines in the top right panel). At the same time >10 S/N Ly$\alpha$ autopower and 21cm cross-power measurements could already be feasible for lower spectral resolution probes like the approved SPHEREx mission~\citep{2014spherex} if sufficient sensitivities of $\sigma_\mathrm{N}\sim 10^{-18}$erg s$^{-1}$cm$^{-2}$sr$^{-1}$Hz$^{-1}$ are reached (see the dashed lines in the top right panel). \rev{Note that a sensitivity of $\sigma_\mathrm{N}\sim 10^{-18}$erg s$^{-1}$cm$^{-2}$sr$^{-1}$Hz$^{-1}$ (conservatively) assumes the minimum science requirement of the SPHEREx all-sky survey. For comparison, for Ly$\alpha$ at $z=8$ the current best estimate is located at $\sigma_\mathrm{N}\sim 8\times 10^{-20}$erg s$^{-1}$cm$^{-2}$sr$^{-1}$Hz$^{-1}$ (dotted lines labelled 'SPHEREx');\footnote{See public sensitivity estimates for SPHEREx here: https://github.com/SPHEREx/Public-products/} we also display an estimate for optimised sensitivity in reach for a CDIM-type mission at $\sigma_\mathrm{N}\sim 1.5\times 10^{-21}$erg s$^{-1}$cm$^{-2}$sr$^{-1}$Hz$^{-1}$ (dotted lines labelled 'CDIM'). In Figure~\ref{fig:SN}, right panels, we perturb around these $\sigma_\mathrm{N}$ values to cover the full range of possible $\sigma_\mathrm{N}$.} We here assumed for 21cm foreground treatment that modes in \rev{an extended} foreground wedge are discarded \rev{in the moderate 'mod' scenario that also encompasses the presence of spectrally smooth foregrounds in our k-range of interest. Note though that in this foreground scenario for sensitivities below $\sigma_\mathrm{N}\sim 10^{-18}$erg s$^{-1}$cm$^{-2}$sr$^{-1}$Hz$^{-1}$ a SPHEREx-type survey will not reach S/N>1 at $z\sim8$.} If instead for comparison the 21cm signal can be modelled well into the wedge region (i.e. the EoR region expanded and foregrounds limited to the primary field of view of the 21cm experiment), both autopower and cross-power measurements are possible down to sensitivities of a few $\sigma_\mathrm{N}\sim 10^{-17}$erg s$^{-1}$cm$^{-2}$sr$^{-1}$Hz$^{-1}$  (see the broken off \rev{orange} drawn and dashed lines in the right panels, labelled 'opt'). \rev{We find the 'opt' foreground scenario improves the cumulative S/N up to a factor of $\sim4$ across survey setups tested, especially towards low sensitivities. For comparison the bottom right panel shows the cumulative S/N of the cross-power with 21cm as a function of sensitivity for a HERA-like setup for both CDIM-type and SPEHREx-type missions, as well as 'mod' (cyan) and 'opt' (orange) foreground treatments. It can be seen that already a cross-signal detection between HERA and SPHEREx is in reach in the case of optimal 'opt' foreground treatment that manages to restrict foregrounds mostly to the primary  field of view of the 21cm instrument.}

Concerning the detector pixel size, between $x_\mathrm{pix}\sim1-8\,$arcsec, the S/N-curve for the cross-power spectrum of Ly$\alpha$ as a representative line for multi-line IM and 21cm is relatively flat, pointing towards a relatively mild requirement in terms of resolution in transverse modes $k_\mathrm{\perp, res}$. For the autopower in Ly$\alpha$ the S/N rises though to $>10^3$ per-redshift bin for pixel sizes of $\sim 1\,$arcsec, which stresses the value of additional transverse resolution for intensity mapping line surveys on their own.

Similarly, we show in Figure~\ref{fig:SNHa} the cumulative S/N of both H$\alpha$ autopower and cross-power spectra with 21cm, for CDIM-type (drawn lines) and SPHEREx-type (dashed lines) experiments. It is important to note though, that the wavelength range of SPHEREx extends only up to 5 microns and thus does not cover H$\alpha$ for $z>5$. While for H$\alpha$ alone the S/N is decreased by a factor of 2-3 in our fiducial setup of a CDIM-type experiments at high sensitivities $\sigma_\mathrm{N}$, at lower line sensitivities the S/N remains higher. Also at lower sensitivities the cross-power with 21cm displays higher S/N as compared to Ly$\alpha$. We note, that the S/N for H$\alpha$ auto and cross signals is only mildly dependent on the spectral resolution and pixel sizes probed in this study, with S/N values of similar magnitude as observed for Ly$\alpha$. In addition we note that in the presence of foregrounds and limited instrumental resolution as investigated here, survey areas starting from a CDIM-like deep, medium, or wide setup ($\sim$ 15, 31, 100 deg$^2$) do not significantly impact the cumulative signal-to-noise attainable, \rev{at least for scales k larger than a few 0.01$\,$Mpc$^{-1}$ investigated here}.

\subsubsection*{\rev{Requirements and k-space overlap}}

We finish by stressing, that on their own, multi-line IM surveys reach signal-to-noise values for tomographic imaging finely spaced in redshift far into the epoch of reionization for moderate spectral resolution of O(100), moderate transverse resolution and sensitivities of $10^{-18}$ to $10^{-19}$erg s$^{-1}$cm$^{-2}$sr$^{-1}$Hz$^{-1}$. To optimally exploit synergies with the 21cm signal to map out the process of reionization and learn about source properties, a spectral resolution $R>100$ is advantageous. This is due to the window where measurable transverse and parallel modes overlap for the 21cm signal versus multi-line IM. This becomes obvious in Figure~\ref{fig:SN-wedge}, where the 2D cylindrically-averaged power spectrum for Ly$\alpha$ at $z\sim 8$ is shown together with the wedge-cut for 21cm foregrounds and horizontal limits in parallel modes induced by the spectral resolution. The 'wedge' of cross-detection is confined to the area below the horizontal limits outside the 21cm-wedge, and becomes significantly narrowed down for values below $R\sim 100$.

\section{Cookbook for successful multi-line IM}\label{sec:cookbook}
We briefly summarise our main findings and recommendations for optimal synergies in multi-line IM, based on our study of Ly$\alpha$, H$\alpha$ and 21cm autopower and cross-power spectra and derived sensitivities in the previous sections. Our analysis for successful multi-line IM finds:
\begin{itemize}
    \item  High S/N measurements of O($10^3$) per z-bin over several orders in scale throughout the EoR are in reach for our fiducial mission setups, pushing EoR observations to higher redshifts of $z\sim8-10$ than for example accessible to LAE surveys;
    \item For IM autopower, spectral resolutions of just R$\sim$O(10), line sensitivities above $\sim 10^{-18}$ erg s$^{-1}$cm$^{-2}$sr$^{-1}$Hz$^{-1}$ and spatial pixel sizes below $\sim 10\,$arcsec are sufficient for high (>100) S/N measurements at redshifts above 7;
    \item To reach a plateau of S/N>1000 per z-bin in IM autopower only moderately increased requirements of R$>250$, sensitivity $>8\times 10^{-18}$ erg s$^{-1}$cm$^{-2}$sr$^{-1}$Hz$^{-1}$ and pixel sizes $<2\,$arcsec are needed;
    \item Due to the milder S/N dependency on instrument characteristics for H$\alpha$ as opposed to Ly$\alpha$, requirements are driven by Ly$\alpha$ IM;
    \item For better k-space overlap with 21cm experiments (due to large-scale modes lost to foregrounds), a spectral resolution of minimum R$\sim$O(100) is key;
    \item Lower resolution missions such as SPHEREx already can provide sufficient S/N in the cross-signal if the 21cm EoR window can be pushed into the foreground wedge region, \rev{both for an SKA1-Low and HERA-type experiment};
    \item Survey areas larger than $\sim 15$deg$^2$ for our fiducial IM mission did not significantly increase the derived cumulative S/N at scales investigated.
\end{itemize}
Missions designed in this flavor of multi-line IM will, with moderate requirements on e.g. the spectral resolution for a  <1m-class instrument in space, be able to closely follow the progress of reionization, supply additional information on the sources that drive reionization, and serve as a valuable cross-check for the cosmological 21cm signal. As this analysis was based on assessing the detectability of autopower and cross-power spectra for optimised multi-line IM during the EoR, depending on question asked and model, other requirements might necessitate a different set of recommendations. For example when aiming for large-scale cosmological effects close to the horizon, increased survey sizes comes with increased information gain on the underlying density field and cosmology at these scales. 

\section{Conclusions and Outlook}\label{sec:out}
In this work we have employed self-consistent simulations of 21cm, Ly$\alpha$ and H$\alpha$ intensity maps to explore optimal survey parameters for both a multi-line IM mission on its own and for synergies with measurements of the 21cm line during the epoch of reionization. To judge optimal information gain, we focused on the cumulative signal-to-noise per tomographic slice attainable for both autopower and cross-power statistics. 

We find that for moderate requirements in terms of spectral resolution, line sensitivity and spatial pixel size, as well as for survey areas moderately-sized for intensity mapping wide-field missions, high signal-to-noise measurements per tomographic redshift bin and for a large range in redshift are able to cover the epoch of reionization. To maximise overlap in terms of accessible modes shared between IR multi-line IM missions and 21cm IM surveys as performed for example by SKA1-Low, for redshifts z>7 a spectral resolution of R>100 is advantageous. At the same time, while the 21cm foreground window restricts accessible shared modes, low-redshift interloping lines do not pose a problem for multi-line IM, as they can be largely projected out by low-level mitigation such as high flux masking. 

Turning the 'problem' of missing overlap between IR and 21cm IM surveys due to foreground cuts into an asset, large scale modes lost to bright foregrounds in the so-called wedge for interferometric 21cm observations can partially be recovered when combining them with photometric and spectroscopic galaxy survey data, see also~\citet{Modi2021}. Equally important, more sophisticated active modelling of 21cm foregrounds well into the wedge-region can be crucial to increase overlap with low-resolution wide-field multi-line IM missions during the EoR. We have shown, that already the approved SPHEREx mission will be able to be cross-correlated with 21cm measurements for high S/N measurements in an optimistic foreground scenario. 

High signal-to-noise tomographic intensity mappings of multiple lines throughout reionization herald a multitude of possibilities. Besides following the global process of reionization and growth of ionized regions directly via imaging as well as via cross-statistics as the 21cm -- Ly$\alpha$ cross-correlation, global properties of ionising sources such as their typical escape fraction and fraction of galactic versus diffuse components in emission, their luminosity function, and typical gas properties can be derived from these maps.  In upcoming studies we will strive to explore some of these synergies and measurable model properties such as the escape fraction of ionising radiation. Lastly, the cross-correlation of several lines, especially for high-redshift measurements, is less prone to systematics and foreground contamination, and therefore is one possible smoking-gun for the cosmological nature of the 21cm signal, as is the cross-correlation with e.g. Lyman-alpha emitting galaxies towards the later stages of reionization~\citep{Heneka2020}.

\section*{Acknowledgements}

We would like to thank the anonymous referee for the useful and constructive suggestions that helped to improve this paper. We thank Guochao Sun for useful feedback and suggestions on the draft. CH would like to thank Andrei Mesinger for useful discussions in the course of this project. 
CH acknowledges support by the Deutsche Forschungsgemeinschaft (DFG, German Research Foundation) under Germany’s Excellence Strategy – EXC 2121 „Quantum Universe“ – 390833306." AC acknowledges support from
80NSSC20K0437 and 80NSSC20K1247.

\section*{Data Availability}

The simulation data and sensitivity curves underlying this article will be shared on reasonable request to the corresponding author.




\bibliographystyle{mnras}
\bibliography{IM_survey} 

\begin{thebibliography}{}
\makeatletter
\relax
\def\mn@urlcharsother{\let\do\@makeother \do\$\do\&\do\#\do\^\do\_\do\%\do\~}
\def\mn@doi{\begingroup\mn@urlcharsother \@ifnextchar [ {\mn@doi@}
  {\mn@doi@[]}}
\def\mn@doi@[#1]#2{\def\@tempa{#1}\ifx\@tempa\@empty \href
  {http://dx.doi.org/#2} {doi:#2}\else \href {http://dx.doi.org/#2} {#1}\fi
  \endgroup}
\def\mn@eprint#1#2{\mn@eprint@#1:#2::\@nil}
\def\mn@eprint@arXiv#1{\href {http://arxiv.org/abs/#1} {{\tt arXiv:#1}}}
\def\mn@eprint@dblp#1{\href {http://dblp.uni-trier.de/rec/bibtex/#1.xml}
  {dblp:#1}}
\def\mn@eprint@#1:#2:#3:#4\@nil{\def\@tempa {#1}\def\@tempb {#2}\def\@tempc
  {#3}\ifx \@tempc \@empty \let \@tempc \@tempb \let \@tempb \@tempa \fi \ifx
  \@tempb \@empty \def\@tempb {arXiv}\fi \@ifundefined
  {mn@eprint@\@tempb}{\@tempb:\@tempc}{\expandafter \expandafter \csname
  mn@eprint@\@tempb\endcsname \expandafter{\@tempc}}}

\bibitem[\protect\citeauthoryear{Barkana \& Loeb}{Barkana \&
  Loeb}{2001}]{Barkana:2000fd}
Barkana R.,  Loeb A.,  2001, \mn@doi [Phys. Rept.]
  {10.1016/S0370-1573(01)00019-9}, 349, 125

\bibitem[\protect\citeauthoryear{Beardsley et~al.}{Beardsley
  et~al.}{2016}]{Beardsley:2016njr}
Beardsley A.~P.,  et~al., 2016, \mn@doi [Astrophys. J.]
  {10.3847/1538-4357/833/1/102}, 833, 102

\bibitem[\protect\citeauthoryear{{Brax}, {Clesse}  \& {Davis}}{{Brax}
  et~al.}{2013}]{2013JCAP...01..003B}
{Brax} P.,  {Clesse} S.,   {Davis} A.-C.,  2013, \mn@doi [\jcap]
  {10.1088/1475-7516/2013/01/003}, \href
  {https://ui.adsabs.harvard.edu/abs/2013JCAP...01..003B} {2013, 003}

\bibitem[\protect\citeauthoryear{{Cooray} et~al.,}{{Cooray}
  et~al.}{2016}]{2016arXiv160205178C}
{Cooray} A.,  et~al., 2016, arXiv e-prints, \href
  {https://ui.adsabs.harvard.edu/abs/2016arXiv160205178C} {p. arXiv:1602.05178}

\bibitem[\protect\citeauthoryear{{Cooray} et~al.,}{{Cooray}
  et~al.}{2019a}]{CDIMreport}
{Cooray} A.,  et~al., 2019a, in Bulletin of the American Astronomical Society.
  p.~23 (\mn@eprint {arXiv} {1903.03144})

\bibitem[\protect\citeauthoryear{{Cooray} et~al.,}{{Cooray}
  et~al.}{2019b}]{2019BAAS...51g..23Cf}
{Cooray} A.,  et~al., 2019b, in Bulletin of the American Astronomical Society.
  p.~23 (\mn@eprint {arXiv} {1903.03144})

\bibitem[\protect\citeauthoryear{Crites et~al.,}{Crites
  et~al.}{2014}]{10.1117/12.2057207}
Crites A.~T.,  et~al., 2014, in Holland W.~S.,  Zmuidzinas J.,  eds,
  International Society for Optics and Photonics Vol. 9153, Millimeter,
  Submillimeter, and Far-Infrared Detectors and Instrumentation for Astronomy
  VII. SPIE, pp 613 -- 621, \mn@doi{10.1117/12.2057207}, \url
  {https://doi.org/10.1117/12.2057207}

\bibitem[\protect\citeauthoryear{{Dayal} \& {Ferrara}}{{Dayal} \&
  {Ferrara}}{2018}]{2018revPratika}
{Dayal} P.,  {Ferrara} A.,  2018, \mn@doi [\physrep]
  {10.1016/j.physrep.2018.10.002}, \href
  {https://ui.adsabs.harvard.edu/abs/2018PhR...780....1D} {780, 1}

\bibitem[\protect\citeauthoryear{DeBoer et~al.}{DeBoer
  et~al.}{2016}]{DeBoer:2016tnn}
DeBoer D.~R.,  et~al., 2016, preprint, \href
  {http://adsabs.harvard.edu/abs/2016arXiv160607473D} {} (\mn@eprint {arXiv}
  {1606.07473})

\bibitem[\protect\citeauthoryear{{Dillon} et~al.,}{{Dillon}
  et~al.}{2014}]{Dillon14}
{Dillon} J.~S.,  et~al., 2014, \mn@doi [\prd] {10.1103/PhysRevD.89.023002},
  \href {https://ui.adsabs.harvard.edu/abs/2014PhRvD..89b3002D} {89, 023002}

\bibitem[\protect\citeauthoryear{{Dor{\'e}} et~al.,}{{Dor{\'e}}
  et~al.}{2014}]{2014spherex}
{Dor{\'e}} O.,  et~al., 2014, preprint, \href
  {http://adsabs.harvard.edu/abs/2014arXiv1412.4872D} {} (\mn@eprint {arXiv}
  {1412.4872})

\bibitem[\protect\citeauthoryear{Dumitru, Kulkarni, Lagache  \&
  Haehnelt}{Dumitru et~al.}{2019}]{10.1093/mnras/stz617}
Dumitru S.,  Kulkarni G.,  Lagache G.,   Haehnelt M.~G.,  2019, \mn@doi
  [Monthly Notices of the Royal Astronomical Society] {10.1093/mnras/stz617},
  485, 3486

\bibitem[\protect\citeauthoryear{Feng, Di-Matteo, Croft, Bird, Battaglia  \&
  Wilkins}{Feng et~al.}{2015}]{BlueTides15}
Feng Y.,  Di-Matteo T.,  Croft R.~A.,  Bird S.,  Battaglia N.,   Wilkins S.,
  2015, \mn@doi [Monthly Notices of the Royal Astronomical Society]
  {10.1093/mnras/stv2484}, 455, 2778

\bibitem[\protect\citeauthoryear{{Field}}{{Field}}{1958}]{1958PIRE...46..240F}
{Field} G.~B.,  1958, \mn@doi [Proceedings of the IRE]
  {10.1109/JRPROC.1958.286741}, \href
  {http://adsabs.harvard.edu/abs/1958PIRE...46..240F} {46, 240}

\bibitem[\protect\citeauthoryear{Fonseca, Silva, Santos  \& Cooray}{Fonseca
  et~al.}{2016}]{10.1093/mnras/stw2470}
Fonseca J.,  Silva M.~B.,  Santos M.~G.,   Cooray A.,  2016, \mn@doi [Monthly
  Notices of the Royal Astronomical Society] {10.1093/mnras/stw2470}, 464, 1948

\bibitem[\protect\citeauthoryear{{Furlanetto} \& {Lidz}}{{Furlanetto} \&
  {Lidz}}{2007}]{2007ApJ...660.1030F}
{Furlanetto} S.~R.,  {Lidz} A.,  2007, \mn@doi [Astrophys. J.]
  {10.1086/513009}, \href {http://adsabs.harvard.edu/abs/2007ApJ...660.1030F}
  {660, 1030}

\bibitem[\protect\citeauthoryear{{Gong}, {Cooray}, {Silva}, {Zemcov}, {Feng},
  {Santos}, {Dore}  \& {Chen}}{{Gong} et~al.}{2017}]{2017ApJ...835..273G}
{Gong} Y.,  {Cooray} A.,  {Silva} M.~B.,  {Zemcov} M.,  {Feng} C.,  {Santos}
  M.~G.,  {Dore} O.,   {Chen} X.,  2017, \mn@doi [\apj]
  {10.3847/1538-4357/835/2/273}, \href
  {https://ui.adsabs.harvard.edu/abs/2017ApJ...835..273G} {835, 273}

\bibitem[\protect\citeauthoryear{{Gunn} \& {Peterson}}{{Gunn} \&
  {Peterson}}{1965}]{1965GP}
{Gunn} J.~E.,  {Peterson} B.~A.,  1965, \mn@doi [Astrophys. J.]
  {10.1086/148444}, \href {http://adsabs.harvard.edu/abs/1965ApJ...142.1633G}
  {142, 1633}

\bibitem[\protect\citeauthoryear{{Hayes}, {Schaerer}, {{\"O}stlin},
  {Mas-Hesse}, {Atek}  \& {Kunth}}{{Hayes} et~al.}{2011}]{Hayes:2011}
{Hayes} M.,  {Schaerer} D.,  {{\"O}stlin} G.,  {Mas-Hesse} J.~M.,  {Atek} H.,
  {Kunth} D.,  2011, \mn@doi [Astrophys. J.] {10.1088/0004-637X/730/1/8}, \href
  {http://adsabs.harvard.edu/abs/2011ApJ...730....8H} {730, 8}

\bibitem[\protect\citeauthoryear{{Heneka} \& {Amendola}}{{Heneka} \&
  {Amendola}}{2018}]{2018JCAP...10..004H}
{Heneka} C.,  {Amendola} L.,  2018, \mn@doi [\jcap]
  {10.1088/1475-7516/2018/10/004}, \href
  {https://ui.adsabs.harvard.edu/abs/2018JCAP...10..004H} {2018, 004}

\bibitem[\protect\citeauthoryear{{Heneka} \& {Mesinger}}{{Heneka} \&
  {Mesinger}}{2020}]{Heneka2020}
{Heneka} C.,  {Mesinger} A.,  2020, \mn@doi [\mnras] {10.1093/mnras/staa1517},
  \href {https://ui.adsabs.harvard.edu/abs/2020MNRAS.496..581H} {496, 581}

\bibitem[\protect\citeauthoryear{{Heneka}, {Cooray}  \& {Feng}}{{Heneka}
  et~al.}{2017}]{Heneka17}
{Heneka} C.,  {Cooray} A.,   {Feng} C.,  2017, \mn@doi [\apj]
  {10.3847/1538-4357/aa8eed}, \href
  {https://ui.adsabs.harvard.edu/abs/2017ApJ...848...52H} {848, 52}

\bibitem[\protect\citeauthoryear{{Hutter}, {Dayal}, {M{\"u}ller}  \&
  {Trott}}{{Hutter} et~al.}{2017}]{Hutter:2016}
{Hutter} A.,  {Dayal} P.,  {M{\"u}ller} V.,   {Trott} C.~M.,  2017, \mn@doi
  [\apj] {10.3847/1538-4357/836/2/176}, \href
  {https://ui.adsabs.harvard.edu/abs/2017ApJ...836..176H} {836, 176}

\bibitem[\protect\citeauthoryear{{Hutter}, {Dayal}, {Yepes}, {Gottl{\"o}ber},
  {Legrand}  \& {Ucci}}{{Hutter} et~al.}{2020}]{Astra2020}
{Hutter} A.,  {Dayal} P.,  {Yepes} G.,  {Gottl{\"o}ber} S.,  {Legrand} L.,
  {Ucci} G.,  2020, arXiv e-prints, \href
  {https://ui.adsabs.harvard.edu/abs/2020arXiv200408401H} {p. arXiv:2004.08401}

\bibitem[\protect\citeauthoryear{Ihle et~al.,}{Ihle et~al.}{2019}]{Ihle_2019}
Ihle H.~T.,  et~al., 2019, \mn@doi [The Astrophysical Journal]
  {10.3847/1538-4357/aaf4bc}, 871, 75

\bibitem[\protect\citeauthoryear{Kennicutt}{Kennicutt}{1998}]{Kennicutt:1997ng}
Kennicutt Jr. R.~C.,  1998, \mn@doi [Astrophys. J.] {10.1086/305588}, 498, 541

\bibitem[\protect\citeauthoryear{{Kubota}, {Inoue}, {Hasegawa}  \&
  {Takahashi}}{{Kubota} et~al.}{2020}]{Kubota20}
{Kubota} K.,  {Inoue} A.~K.,  {Hasegawa} K.,   {Takahashi} K.,  2020, \mn@doi
  [\mnras] {10.1093/mnras/staa979}, \href
  {https://ui.adsabs.harvard.edu/abs/2020MNRAS.494.3131K} {494, 3131}

\bibitem[\protect\citeauthoryear{Li, Wechsler, Devaraj  \& Church}{Li
  et~al.}{2016}]{Li_2016}
Li T.~Y.,  Wechsler R.~H.,  Devaraj K.,   Church S.~E.,  2016, \mn@doi [The
  Astrophysical Journal] {10.3847/0004-637x/817/2/169}, 817, 169

\bibitem[\protect\citeauthoryear{{Lidz}, {Zahn}, {McQuinn}, {Zaldarriaga}  \&
  {Hernquist}}{{Lidz} et~al.}{2008}]{2008LidzNoise}
{Lidz} A.,  {Zahn} O.,  {McQuinn} M.,  {Zaldarriaga} M.,   {Hernquist} L.,
  2008, \mn@doi [Astrophys. J.] {10.1086/587618}, \href
  {http://adsabs.harvard.edu/abs/2008ApJ...680..962L} {680, 962}

\bibitem[\protect\citeauthoryear{{Lidz}, {Furlanetto}, {Oh}, {Aguirre},
  {Chang}, {Dor{\'e}}  \& {Pritchard}}{{Lidz}
  et~al.}{2011}]{2011ApJ...741...70L}
{Lidz} A.,  {Furlanetto} S.~R.,  {Oh} S.~P.,  {Aguirre} J.,  {Chang} T.-C.,
  {Dor{\'e}} O.,   {Pritchard} J.~R.,  2011, \mn@doi [Astrophys. J.]
  {10.1088/0004-637X/741/2/70}, \href
  {http://adsabs.harvard.edu/abs/2011ApJ...741...70L} {741, 70}

\bibitem[\protect\citeauthoryear{{Liu}, {Heneka}  \& {Amendola}}{{Liu}
  et~al.}{2020}]{2020JCAP...05..038L}
{Liu} X.-W.,  {Heneka} C.,   {Amendola} L.,  2020, \mn@doi [\jcap]
  {10.1088/1475-7516/2020/05/038}, \href
  {https://ui.adsabs.harvard.edu/abs/2020JCAP...05..038L} {2020, 038}

\bibitem[\protect\citeauthoryear{McAlpine et~al.,}{McAlpine
  et~al.}{2016}]{MCALPINE201672}
McAlpine S.,  et~al., 2016, \mn@doi [Astronomy and Computing]
  {https://doi.org/10.1016/j.ascom.2016.02.004}, 15, 72

\bibitem[\protect\citeauthoryear{{Mesinger} \& {Furlanetto}}{{Mesinger} \&
  {Furlanetto}}{2007}]{DexM07}
{Mesinger} A.,  {Furlanetto} S.,  2007, \mn@doi [Astrophys. J.]
  {10.1086/521806}, \href {http://adsabs.harvard.edu/abs/2007ApJ...669..663M}
  {669, 663}

\bibitem[\protect\citeauthoryear{{Mesinger}, {Furlanetto}  \& {Cen}}{{Mesinger}
  et~al.}{2011}]{Mesinger10}
{Mesinger} A.,  {Furlanetto} S.,   {Cen} R.,  2011, \mn@doi [MNRAS]
  {10.1111/j.1365-2966.2010.17731.x}, \href
  {http://adsabs.harvard.edu/abs/2011MNRAS.411..955M} {411, 955}

\bibitem[\protect\citeauthoryear{Miralda-Escud{\'{e}}}{Miralda-Escud{\'{e}}}{1998}]{Escude98}
Miralda-Escud{\'{e}} J.,  1998, The Astrophysical Journal, 501, 15

\bibitem[\protect\citeauthoryear{{Modi}, {White}, {Castorina}  \&
  {Slosar}}{{Modi} et~al.}{2021}]{Modi2021}
{Modi} C.,  {White} M.,  {Castorina} E.,   {Slosar} A.,  2021, arXiv e-prints,
  \href {https://ui.adsabs.harvard.edu/abs/2021arXiv210208116M} {p.
  arXiv:2102.08116}

\bibitem[\protect\citeauthoryear{{Nelson} et~al.,}{{Nelson}
  et~al.}{2019}]{TNG2019}
{Nelson} D.,  et~al., 2019, \mn@doi [Computational Astrophysics and Cosmology]
  {10.1186/s40668-019-0028-x}, \href
  {https://ui.adsabs.harvard.edu/abs/2019ComAC...6....2N} {6, 2}

\bibitem[\protect\citeauthoryear{Ouchi et~al.,}{Ouchi
  et~al.}{2017}]{10.1093/pasj/psx074}
Ouchi M.,  et~al., 2017, \mn@doi [Publications of the Astronomical Society of
  Japan] {10.1093/pasj/psx074}, 70

\bibitem[\protect\citeauthoryear{{Ouchi}, {Ono}  \& {Shibuya}}{{Ouchi}
  et~al.}{2020}]{2020ARA&A..58..617O}
{Ouchi} M.,  {Ono} Y.,   {Shibuya} T.,  2020, \mn@doi [\araa]
  {10.1146/annurev-astro-032620-021859}, \href
  {https://ui.adsabs.harvard.edu/abs/2020ARA&A..58..617O} {58, 617}

\bibitem[\protect\citeauthoryear{{Pober} et~al.,}{{Pober}
  et~al.}{2013}]{Pober:2013}
{Pober} J.~C.,  et~al., 2013, \mn@doi [\aj] {10.1088/0004-6256/145/3/65}, \href
  {https://ui.adsabs.harvard.edu/abs/2013AJ....145...65P} {145, 65}

\bibitem[\protect\citeauthoryear{Pober et~al.}{Pober
  et~al.}{2014}]{Pober:2013jna}
Pober J.~C.,  et~al., 2014, \mn@doi [Astrophys. J.]
  {10.1088/0004-637X/782/2/66, 10.1088/0004-637X/788/1/96}, 782, 66

\bibitem[\protect\citeauthoryear{{Qin}, {Duffy}, {Mutch}, {Poole}, {Mesinger}
  \& {Wyithe}}{{Qin} et~al.}{2019}]{2019MNRAS.487.1946Q}
{Qin} Y.,  {Duffy} A.~R.,  {Mutch} S.~J.,  {Poole} G.~B.,  {Mesinger} A.,
  {Wyithe} J. S.~B.,  2019, \mn@doi [\mnras] {10.1093/mnras/stz1380}, \href
  {https://ui.adsabs.harvard.edu/abs/2019MNRAS.487.1946Q} {487, 1946}

\bibitem[\protect\citeauthoryear{{Razoumov} \& {Sommer-Larsen}}{{Razoumov} \&
  {Sommer-Larsen}}{2010}]{Razoumov:2010}
{Razoumov} A.~O.,  {Sommer-Larsen} J.,  2010, \mn@doi [Astrophys. J.]
  {10.1088/0004-637X/710/2/1239}, \href
  {http://adsabs.harvard.edu/abs/2010ApJ...710.1239R} {710, 1239}

\bibitem[\protect\citeauthoryear{{Serra}, {Dor{\'e}}  \& {Lagache}}{{Serra}
  et~al.}{2016}]{Serra:2016jzs}
{Serra} P.,  {Dor{\'e}} O.,   {Lagache} G.,  2016, \mn@doi [\apj]
  {10.3847/1538-4357/833/2/153}, \href
  {https://ui.adsabs.harvard.edu/abs/2016ApJ...833..153S} {833, 153}

\bibitem[\protect\citeauthoryear{{Silva}, {Santos}, {Gong}, {Cooray}  \&
  {Bock}}{{Silva} et~al.}{2013}]{Silva12}
{Silva} M.~B.,  {Santos} M.~G.,  {Gong} Y.,  {Cooray} A.,   {Bock} J.,  2013,
  \mn@doi [Astrophys. J.] {10.1088/0004-637X/763/2/132}, \href
  {http://adsabs.harvard.edu/abs/2013ApJ...763..132S} {763, 132}

\bibitem[\protect\citeauthoryear{Silva, Santos, Cooray  \& Gong}{Silva
  et~al.}{2015}]{Silva_2015}
Silva M.,  Santos M.~G.,  Cooray A.,   Gong Y.,  2015, \mn@doi [The
  Astrophysical Journal] {10.1088/0004-637x/806/2/209}, 806, 209

\bibitem[\protect\citeauthoryear{Sobacchi \& Mesinger}{Sobacchi \&
  Mesinger}{2014}]{Sobacchi:2014rua}
Sobacchi E.,  Mesinger A.,  2014, \mn@doi [Mon. Not. Roy. Astron. Soc.]
  {10.1093/mnras/stu377}, 440, 1662

\bibitem[\protect\citeauthoryear{{Sobacchi}, {Mesinger}  \& {Greig}}{{Sobacchi}
  et~al.}{2016}]{Sobacchi:2016mhx}
{Sobacchi} E.,  {Mesinger} A.,   {Greig} B.,  2016, \mn@doi [\mnras]
  {10.1093/mnras/stw811}, \href
  {https://ui.adsabs.harvard.edu/abs/2016MNRAS.459.2741S} {459, 2741}

\bibitem[\protect\citeauthoryear{{Sun}, {Hensley}, {Chang}, {Dor{\'e}}  \&
  {Serra}}{{Sun} et~al.}{2019}]{Jason2019}
{Sun} G.,  {Hensley} B.~S.,  {Chang} T.-C.,  {Dor{\'e}} O.,   {Serra} P.,
  2019, \mn@doi [\apj] {10.3847/1538-4357/ab55df}, \href
  {https://ui.adsabs.harvard.edu/abs/2019ApJ...887..142S} {887, 142}

\bibitem[\protect\citeauthoryear{Tingay et~al.,}{Tingay
  et~al.}{2013}]{tingay2013}
Tingay S.~J.,  et~al., 2013, \mn@doi [Publications of the Astronomical Society
  of Australia] {10.1017/pasa.2012.007}, 30, e007

\bibitem[\protect\citeauthoryear{Vrbanec, Ciardi, Jelić, Jensen, Iliev,
  Mellema  \& Zaroubi}{Vrbanec et~al.}{2020}]{10.1093/mnras/staa183}
Vrbanec D.,  Ciardi B.,  Jelić V.,  Jensen H.,  Iliev I.~T.,  Mellema G.,
  Zaroubi S.,  2020, \mn@doi [Monthly Notices of the Royal Astronomical
  Society] {10.1093/mnras/staa183}, 492, 4952

\bibitem[\protect\citeauthoryear{{Wouthuysen}}{{Wouthuysen}}{1952}]{1952AJ.....57R..31W}
{Wouthuysen} S.~A.,  1952, \mn@doi [Astron. J.] {10.1086/106661}, \href
  {http://adsabs.harvard.edu/abs/1952AJ.....57R..31W} {57, 31}

\bibitem[\protect\citeauthoryear{{Yang}, {Somerville}, {Pullen}, {Popping},
  {Breysse}  \& {Maniyar}}{{Yang} et~al.}{2020}]{2020arXiv200911933Y}
{Yang} S.,  {Somerville} R.~S.,  {Pullen} A.~R.,  {Popping} G.,  {Breysse}
  P.~C.,   {Maniyar} A.~S.,  2020, arXiv e-prints, \href
  {https://ui.adsabs.harvard.edu/abs/2020arXiv200911933Y} {p. arXiv:2009.11933}

\bibitem[\protect\citeauthoryear{Yue, Ferrara, Pallottini, Gallerani  \&
  Vallini}{Yue et~al.}{2015}]{10.1093/mnras/stv933}
Yue B.,  Ferrara A.,  Pallottini A.,  Gallerani S.,   Vallini L.,  2015,
  \mn@doi [Monthly Notices of the Royal Astronomical Society]
  {10.1093/mnras/stv933}, 450, 3829

\bibitem[\protect\citeauthoryear{{van Haarlem} et~al.}{{van Haarlem}
  et~al.}{2013}]{2013A&A...556A...2V}
{van Haarlem} M.~P.,  et~al., 2013, \mn@doi [Astron. Astrophys.]
  {10.1051/0004-6361/201220873}, \href
  {http://adsabs.harvard.edu/abs/2013A%26A...556A...2V} {556, A2}

\makeatother
\end{thebibliography}



\appendix

\section{L\lowercase{y}$\alpha$ damping}\label{app:damping}

To take into account the resonant nature of the Ly$\alpha$ line, we calculate the optical depth $\tau_\mathrm{Ly\alpha}$ to correct the intrinsic galactic Ly$\alpha$ luminosity. When calculating $\tau_\mathrm{Ly\alpha}$ we trace through the medium along the LOS to account for redshifting between emission and patches of neutral medium around. This means the emission can get shifted from the line core in resonance to the line wings of lower optical depth before reaching surrounding patches of neutral hydrogen. We sum the contributions of neutral patches along the LOS following the analytical result from~\citet{Escude98}, see also~\citet{DexM07},
\begin{align}
\tau_\mathrm{Ly\alpha} \left( z_\mathrm{obs}\right) = &  \tau_\mathrm{s} \sum_i x_{\mathrm{HI},i}  \left( \frac{2.02\times 10^{-8} }{\pi}\right) \left(\frac{1+z_{\mathrm{a}i} }{1+z_\mathrm{obs}}\right)^{1.5} \nonumber \\ &  \vspace{0.2cm}\times \left[ I\left( \frac{1+z_{\mathrm{a}i}}{1+z_\mathrm{obs}}\right) -  I\left( \frac{1+z_{\mathrm{e}i}}{1+z_\mathrm{obs}}\right)\right] , \label{eq:tauLya2}
\end{align} 
where each neutral patch extends from redshift $z_\mathrm{ai}$ to $z_\mathrm{ei}$ with $z_\mathrm{ai}>z_\mathrm{ei}$, $\bar{x}_\mathrm{HI}$ is the average IGM neutral hydrogen fraction, and $\tau_\mathrm{s}$ is the optical depth at line resonance in neutral hydrogen under the assumption of a uniform gas distribution, which can be approximated at high redshifts by~\citep{1965GP,Barkana:2000fd}
\begin{equation}
\tau_\mathrm{s} \approx 6.45\times 10^5 \left( \frac{\Omega_\mathrm{b} h}{0.03}\right) \left( \frac{\Omega_\mathrm{m}}{0.3}\right)^{-0.5} \left( \frac{1+z_\mathrm{s}}{10}\right)^{1.5},
\end{equation}
with source redshift $z_\mathrm{s}$, and present-day density parameters of matter $\Omega_\mathrm{m}$ and of baryons $\Omega_\mathrm{b}$. The helper function $I\left( x\right)$ is defined as
\begin{align}
I\left( x\right) = & \frac{x^{4.5}}{1-x} + \frac{9}{7}x^{3.5} + \frac{9}{5}x^{2.5} + 3x^{1.5} +9x^{0.5 } \nonumber \\ & - 4.5\ln\left( \frac{1+x^{0.5}}{1-x^{0.5}}\right) . 
\end{align}

\section{L\lowercase{y}$\alpha$ $\&$ H$\alpha$ - Power spectra}\label{app:Pks}

In Figure~\ref{app:Pks} we depict the theoretical power spectra for Ly$\alpha$ and H$\alpha$ as derived from our simulations for reference. Note that they represent the unprojected power spectra, not taking into account a specific observational setup. The components shown represent the total emission 'tot', the galactic contribution 'gal', the diffuse IGM contribution 'dIGM' and for Ly$\alpha$ in addition the scattered IGM contribution 'sIGM'. Note for example that H$\alpha$ emission is dominated by the galactic component, while for Ly$\alpha$ emission also the IGM contributions are significant. 

\begin{figure}
\includegraphics[width=\columnwidth]{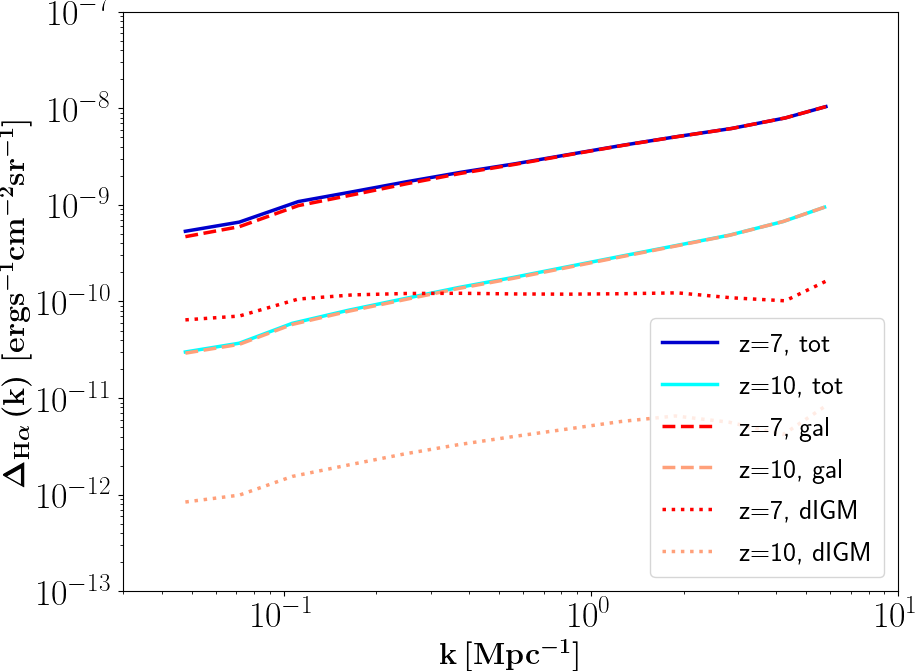}
\includegraphics[width=\columnwidth]{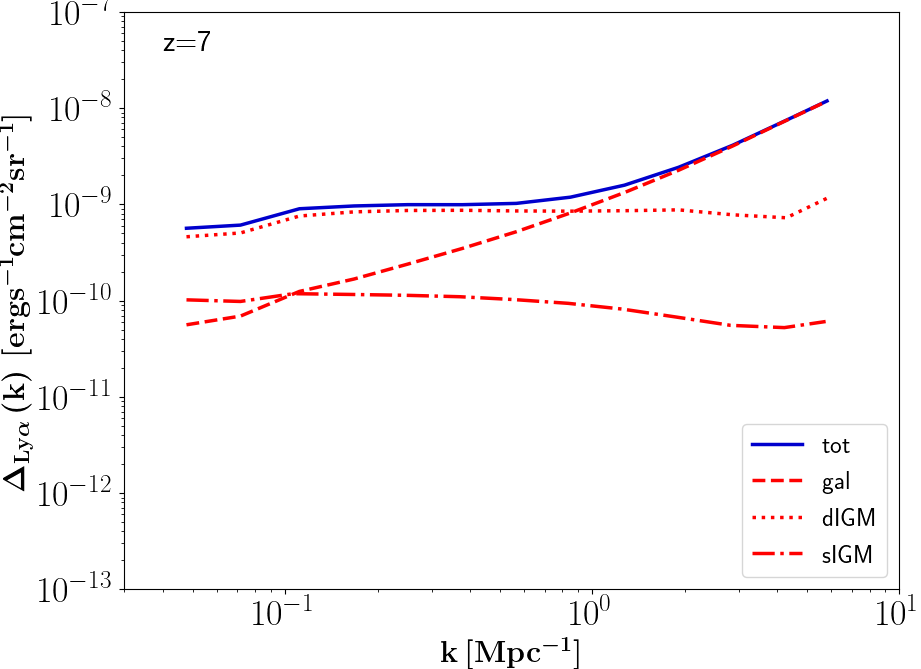}
\includegraphics[width=\columnwidth]{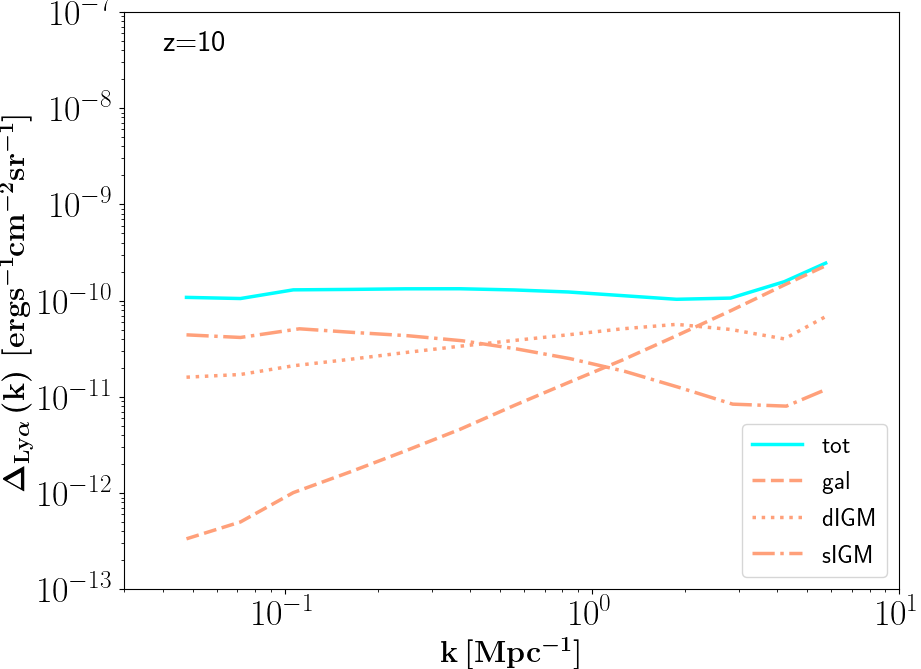}
\caption{H$\alpha$ (top panel) and Ly$\alpha$ (middle panel $z=7$, bottom panel $z=10$) power spectra. Blue (cyan) lines depict the total emission 'tot', dashed red (orange) lines the galactic contribution 'gal', dotted red (orange) lines the diffuse IGM contribution 'dIGM' and dash-dotted red (orange) lines the scattered IGM contribution 'sIGM' at $z=7$ ($z=10$). All power spectra are the non-mission specific unprojected power spectra.}
\label{fig:PksHaLya}
\end{figure}

\section{Comparison of error terms}\label{app:errors}

\begin{figure*}
\centering
\includegraphics[width=\columnwidth]{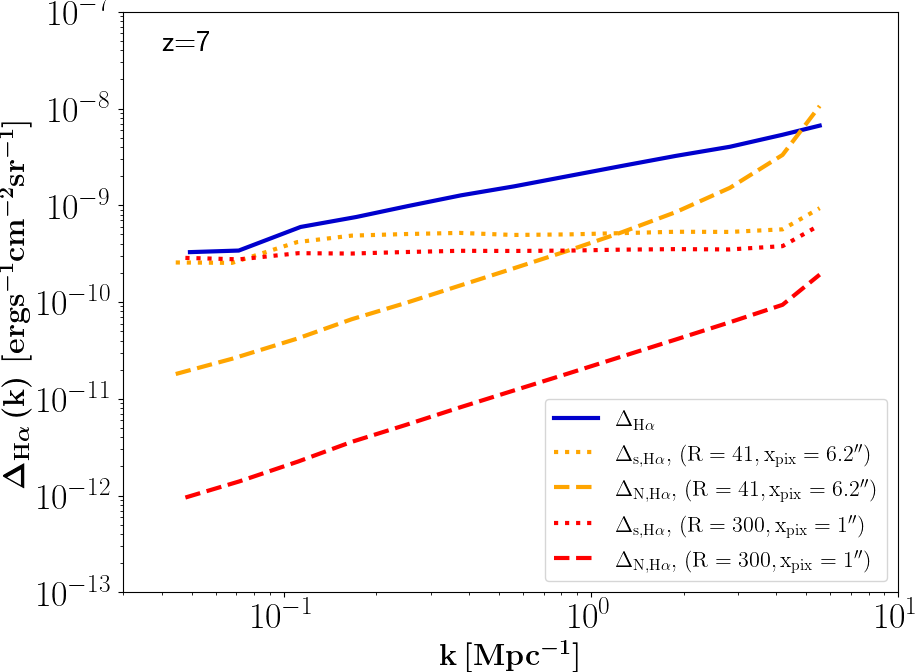}
\includegraphics[width=\columnwidth]{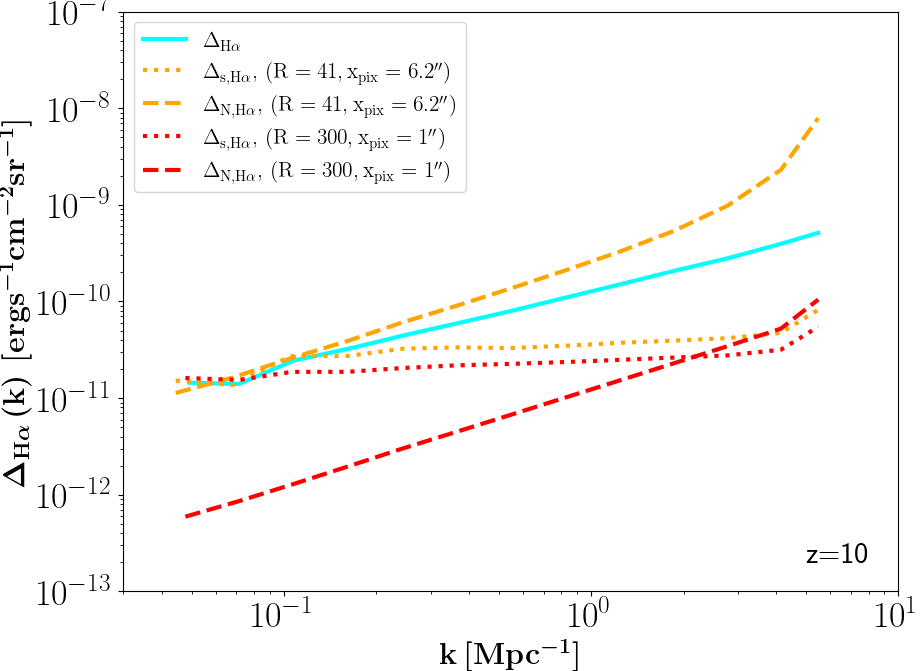}
\includegraphics[width=\columnwidth]{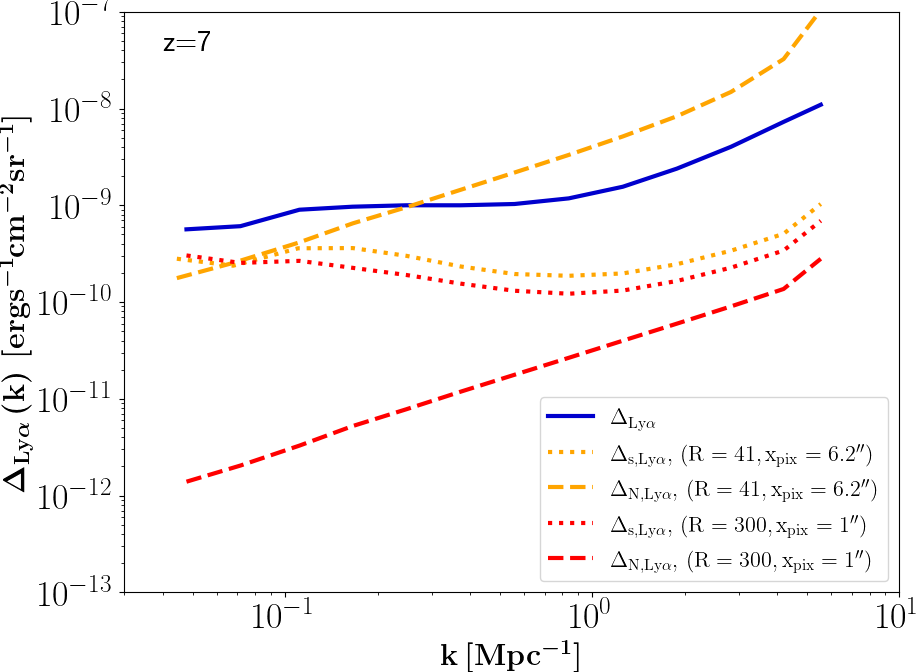}
\includegraphics[width=\columnwidth]{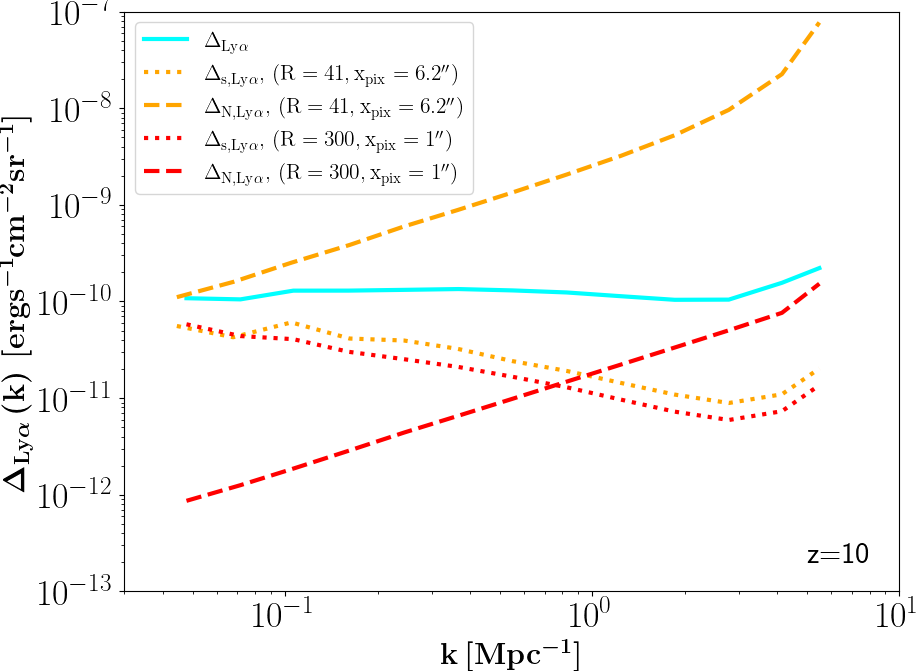}
\caption{Error budget in terms of cosmic variance and noise power at $z=7$ (left panels) and $z=10$ (right panels) during the EoR for two fiducial mission setups, a CDIM-type medium survey (red dashed and dotted lines) and a SPHEREx-type experiment (orange dashed and dotted lines). The blue (cyan) lines show the predicted power spectra at $z=7$ ($z=10$) for H$\alpha$ (top panel) and Ly$\alpha$ (bottom panels). The respective dashed lines represent the instrumental noise power $\Delta_\mathrm{N}$, whereas the dotted lines represent the sample variance contribution to the error budget. Where blue (cyan) lines are above dashed and dotted lines, is the signal expected to be larger than the respective error budget terms. When the noise power (dashed lines) is below / of similar magnitude as the sample variance (dotted lines), the error bars are minimised (the so-called 'Knox criterion'). Note for example that for a CDIM-like survey (red) the error bars are minimised, even at $z=10$, at scales $k\lesssim 1\,$Mpc$^{-1}$ for both H$\alpha$ and Ly$\alpha$.}
\label{fig:errors}
\end{figure*}

The relative importance of sample variance $P_\mathrm{I}$ and noise power spectrum $P_\mathrm{N}$ is an indicator for the best setup that minimises the error bars on the power spectra of interest. For intensity mapping we aim for $P_\mathrm{N} \lesssim P_\mathrm{I}$ for a large range in modes $k$ and redshift $z$. In Figure~\ref{fig:errors} we compare throughout reionisation the two error terms for the Ly$\alpha$ and H$\alpha$ auto power spectra for our two fiducial missions setups, CDIM-type and SPHEREx-type, that we perturb around in this study. The error terms plotted are defined as noise power term $\Delta^2_\mathrm{N,I}=k^3/(2\pi^2V) P_\mathrm{N}$ and sample variance term $\Delta^2_\mathrm{s,I}=k^3/(2\pi^2V) P_\mathrm{I}$ for line $I$. We first note that far a CDIM-type mission even at $z=10$ the error terms stay sub-dominant as compared to the signal both for Ly$\alpha$ and H$\alpha$ measurements. For a SPHEREx-style measurement, H$\alpha$ would be well measurable at $z=7$, but at higher redshifts and even for Ly$\alpha$ at $z=7$ only a tentative detection is possible. Note though that the SPHEREx mission does not cover the wavelength range for H$\alpha$ measurements at $z>5$. To assess if the mission setup minimises error bars, we compare if the noise power (dashed lines) is below and/or of similar magnitude as the sample variance (dotted lines), the so-called 'Knox criterion'. For a CDIM-like survey (red dashed and dotted lines) the error bars are minimised throughout reionisation at scales $k\lesssim 1\,$Mpc$^{-1}$ for both H$\alpha$ and Ly$\alpha$. For a SPHEREx-type mission the Knox criterion is fulfilled e.g. for Ly$\alpha$ at $k\lesssim 0.1\,$Mpc$^{-1}$ and is not reached at higher redshifts (i.e. the error budget is dominated by noise power).

\bsp	
\label{lastpage}
\end{document}